\documentclass{article}

\usepackage{arxiv}

\usepackage[utf8]{inputenc} 
\usepackage[T1]{fontenc}    
\usepackage{hyperref}       
\usepackage{url}            
\usepackage{booktabs}       
\usepackage{amsfonts}       
\usepackage{nicefrac}       
\usepackage{microtype}      
\usepackage{lipsum}
\usepackage{graphicx}
\usepackage{graphicx}
\usepackage{float}
\usepackage{asymptote}
\usepackage{csquotes}
\usepackage{xcolor}
\usepackage{csquotes}
\usepackage{pgf-pie}
\usepackage{makecell}
\usepackage{accents}
\graphicspath{ {./images/} }

\title{LLM-Enabled Multi-Agent Systems: Empirical Evaluation and Insights into Emerging Design Patterns \& Paradigms}

\author{
 Harri Renney \\
  Kaze Technologies\\
  Kaze Consulting\\
  Bath, BA1 2HN, UK \\
  \texttt{harri@kaze-consulting.com} \\
   \And
 Maxim N Nethercott \\
  Kaze Technologies\\
  Kaze Consulting\\
  Bath, BA1 2HN, UK \\
  \texttt{max@kaze-consulting.com} \\
  \And
Nathan Renney \\
  University of the West of England\\
  Bristol, UK \\
  \texttt{nathan.renney@uwe.ac.uk} \\
  \And
  Peter Hayes \\
  Kaze Technologies\\
  Kaze Consulting\\
  Bath, BA1 2HN, UK \\
  \texttt{peter@kaze-consulting.com} \\
}

\begin{document}
\maketitle
\begin{abstract}
This paper formalises the literature on emerging design patterns and paradigms for Large Language Model (LLM)-enabled multi-agent systems (MAS), evaluating their practical utility across various domains. We define key architectural components, including agent orchestration, communication mechanisms, and control-flow strategies, and demonstrate how these enable rapid development of modular, domain-adaptive solutions. Three real-world case studies are tested in controlled, containerised pilots in telecommunications security, national heritage asset management, and utilities customer service automation. Initial empirical results show that, for these case studies, prototypes were delivered within two weeks and pilot-ready solutions within one month, suggesting reduced development overhead compared to conventional approaches and improved user accessibility. However, findings also reinforce limitations documented in the literature, including variability in LLM behaviour that leads to challenges in transitioning from prototype to production maturity. We conclude by outlining critical research directions for improving reliability, scalability, and governance in MAS architectures and the further work needed to mature MAS design patterns to mitigate the inherent challenges.
\end{abstract}


\section{Introduction} \label{sect:s1}

The concept of distributing problem solving across multiple AI programs is a field in contemporary artificial intelligence (AI) that emerged in the late 1970s with works from VR Lesser \cite{10.5555/1624861.1624983} and Hewitt C. \cite{hewitt1977viewing}. Research explored distributed problem solving, in which multiple AI type programs could cooperate to solve complex tasks that were too large for a single monolithic system to solve at the time. Originating from these concepts, the modern terminology around multi-agent systems (MAS) has been formalised \cite{abbas2015organization, dorri2018multi}. With the  recent emergence of a combination of key enabling technologies, including the transformer \cite{vaswani2017attention}, large language models (LLMs) \cite{naveed2025comprehensive}, and access to considerable compute power \cite{sastry2024computing}, previously conceptual multi-agent designs and MAS are now possible to investigate in practice using LLMs \cite{renney2025reimagining}.

Since Google published ``Attention is All you Need'' \cite{vaswani2017attention}, modern language models have been converging on the use of \emph{transformer} deep learning architecture with significant generative results \cite{jai.2025.069226}. When model parameters are scaled up, the word by word predictive mechanism leads to impressive emergent abilities. These include solving certain arithmetic calculations \cite{zhanginterpreting}, creative writing \cite{gomez2023confederacy} and software code generation \cite{fakhoury2024llm}. Scaled up versions of language models like GPT \cite{achiam2023gpt} and PaLM \cite{chowdhery2023palm} are referred to as LLMs, and particular LLMs have been shown to pass professional exams such as the US medical licensing exam \cite{brin2023large} and the US law practise BAR examination \cite{katz2024gpt}, along with convincingly going undetected in writing \cite{wu2025survey, cmc.2025.064747} and code generation \cite{10.1609/aaai.v38i21.30361}.

\begin{figure}[H]
\centering
\includegraphics[scale=0.5]{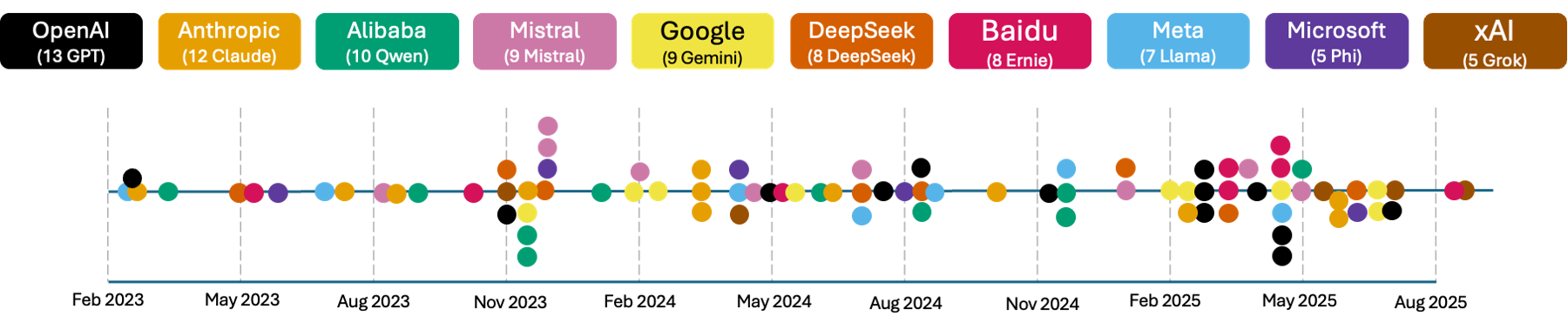}
\caption{Timeline illustrating the cumulative release of major LLM models by ten leading AI companies since 2023 \cite{luyen2025howwell}.}
\label{fig:timeline}
\end{figure}

The prominence of LLMs in contemporary AI research and industry is unmistakable, with leading multinational corporations investing heavily in their development. Individual LLM builds are estimated to cost between \$1 million and \$100 million \cite{chanen2023learnedAI}, and since 2023, at least 86 major releases have been recorded (Figure~\ref{fig:timeline}). This trend reflects an unprecedented acceleration in model innovation and capital expenditure, reinforcing the imperative for scalable and systematic design paradigms to support sustainable deployment.

Using modern LLMs as the core reasoning engine, specialist AI agents can be formed \cite{alto2024building}. With a well-designed prompt, access to up-to-date domain data, and the necessary tools, an agent can be tailored to a specific task or field. The agent then becomes highly specialised for a target task, improving its effectiveness whilst reducing execution time and compute costs \cite{wang2024beyond, kalyuzhnaya2025llm, zhang2025grid}.

By arranging a number of specialist agents in a network that can hand off the next part of the process to another specialised agent, a MAS solution can be formed. Their efficiency stems from the division of labour inherent in MAS, whereby a complex task is divided into multiple smaller tasks, each of which is assigned to a distinct agent \cite{rezaee2015average}. It is this flexibility that as we propose in this paper, makes MAS suited to solve problems in a variety of sectors including utilities and telecoms.

Building on the established foundations of MAS research, this paper articulates the emerging design patterns associated with LLM-enabled MAS and seeks to formalise their use within the community. The approach is evaluated through a series of case studies across multiple domains, from which the results and observations are used to elucidate strengths and limitations. Building on these findings, we identify the critical areas of advancement required to further mature and stabilise this emerging paradigm on the path toward production-ready technology.

\subsection{Paper Structure}

We investigate this space and offer formalisation and results from real-world implementations to better understand the utility and caveats of the LLM-based MAS approach. To that effect, the remainder of this paper is organised as follows:
\begin{itemize}
    \item Section~\ref{sect:litrev} reviews the latest academic, industry and open-source literature on LLM-enabled MAS, concluding with how this paper aims to contribute formalised design patterns and test this approach on real-world case-studies.
    \item Section~\ref{sect:keychallenges} identifies the motivations for using LLM-based MAS solutions, that address key data challenges being faced by organisations.
    \item Section~\ref{sect:paradigm} presents our formalisation of a common MAS design paradigm, outlining how configured specialist agents and network arrangements according to repeatable design patterns can be applied to solve problem classes such as data retrieval and semi-automated, human-in-the-loop pipelines.
    \item Section~\ref{sec:caseseries} documents the application of this approach across three real-world case studies, capturing the experiences and observations of developers, end-users, and senior business executives.
    \item Section~\ref{sec:discuss} provides a discussion of the findings, highlighting both strengths and limitations of the proposed design patterns and broader implications for MAS development.
    \item Finally, Section~\ref{sec:conclusion} concludes the paper with a summary of key insights and future research directions in this evolving field.
\end{itemize}

\section{Literature Review} \label{sect:litrev}

The rapid advances in LLM architectures and the unpredictable emergent behaviours indicate that the technology is still in a formative stage \cite{berti2025emergent}. Exasperated by the lack of understanding of how LLMs achieve their emergent abilities \cite{singh2024rethinking} and ethical concerns leading to potential government regulation \cite{mesko2023imperative}, effective application of LLMs for intelligent agent networks is still ongoing and timely.

One foundational technique that enables LLMs to exhibit improved reasoning capabilities is Chain-of-Thought (CoT) prompting \cite{wei2022chain}. This method guides the model through a series of intermediate reasoning steps before producing a final answer, improving accuracy on tasks requiring logical inference. CoT has proven particularly effective for complex problem-solving and offers an interpretable view of the model’s reasoning process, revealing how conclusions are reached and where errors may arise. While CoT operates sequentially, an extension known as Tree-of-Thought (ToT) introduces branching reasoning paths, yielding measurable improvements on tasks with high complexity \cite{yao2024tree}. Notably, ToT reasoning invokes multiple agents to explore alternative reasoning trajectories, representing a rudimentary form of dynamic multi-agent coordination, a concept that this paper expands upon in greater detail.

Beyond prompt engineering and multi-agent invocation, tool integration has emerged as a critical technique for specialising general-purpose LLMs into task-specific intelligent agents. A prominent example is Retrieval-Augmented Generation (RAG) \cite{lewis2020retrieval}, introduced by Lewis et al. in 2020, which enables LLMs to access and reason over external datasets not included in their training corpus. By indexing dynamic, domain-specific, or proprietary data sources, RAG allows models to generate informed and contextually grounded outputs, making it an industry-standard option for knowledge-intensive applications \cite{oche2025systematic}. As one of RAG’s pioneering researchers notes \cite{Kiela2024RAG}:
\begin{displayquote}
``Language models are about 20\% of the entire system/solution, and is surrounded by further engineering parts.''
\end{displayquote}
emphasising the need to design systems rather than relying on building isolated models. Similarly, other specialised agents require interfaces to interact with diverse tools, such as SQL databases, continuing the evolving drive towards MAS, where multiple LLM-powered specialised agents collaborate to solve complex, heterogeneous tasks.

\subsection{Academic Literature}

As specialised agents proliferate, research increasingly explores how they can be connected and coordinated within MAS to address complex, targetted, domain specific problems. Recent literature presents several MAS designs targeting focused applications. For instance, Okonkwo and Onobhayedo \cite{jai.2024.058649} investigate MAS for requirements capture in software development, demonstrating how LLM-powered agents can assist in eliciting and validating specifications. Similarly, Lin et al. \cite{lin2024llm} propose a MAS framework for code generation, where agents act as domain experts across different stages of the development lifecycle, interacting with engineers in a human–machine teaming environment. Their findings indicate promising efficacy in adopting engineering roles and facilitating collaborative workflows. Beyond software engineering, Kearney’s \emph{Bots of the SoCs} study \cite{kearney2024bots} evaluates MAS integration within Security Operations Centres (SoCs), benchmarking different agent arrangements using the Boss of the SOC dataset \cite{splunkgh2023BotS}. Results suggest that the most effective MAS configuration tested achieves performance comparable to a junior cyber analyst, highlighting MAS potential to augment human expertise in high-stakes domains.

Despite the promising results in the literature, MAS implementations face notable challenges. Pan et al. \cite{pan2025multiagent} highlight that as agent networks scale, coordination complexity can introduce significant overhead, reducing efficiency and responsiveness. This can lead to misalignment of the MAS solutions caused by agents local optimisation conflicting with global goals, the stochastic nature of using LLM's and error propagation where an agent's incorrect reasoning can cascade through a system. Hence, to be effective, the MAS approach needs structured design patterns and paradigms to mitigate intrinsic challenges from MAS including coordination failures and improve reliability. Further, matters of sustainability and ethics will need to be considered as MAS architectures build on LLMs, addressing the emerging challenges inherent to LLM-based systems \cite{ding_sustainable_2024, hinds_our_2023, kapania_im_2025}. Where, in some cases, a prototype may imply the opportunity to defer the importance of issues such as verified accountability, biases, and new vectors of exploitation, it is important that these be considered and addressed \cite{jiao_navigating_2025}. It is therefore important to remain attentive to developments in this area as a better understanding emerges, particularly in areas such as censorship, transparency, and intellectual property and plagiarism, which are currently less explored ethical considerations.

Beyond research focused on domain-specific MAS designs, more ambitious horizon initiatives are emerging. These efforts explore how AI agents might cooperate at scale, potentially evolving into ubiquitous tools that replace existing technologies, including aspects of the internet. A notable example is MIT’s NANDA project \cite{MIT2025NANDA, raskar2025beyond}, which proposes a decentralised \emph{Internet of AI Agents} framework designed to support collaboration among billions of specialised agents across a distributed architecture. Positioned between narrowly focused MAS implementations and visionary concepts of global agent autonomy, this paper focuses on formalising emerging design patterns that enable practical MAS solutions for diverse domain-specific problems.

\subsection{Open-Source \& Industry Initiatives}

Drawing from the activity outside of the academic literature, the open-source community is pushing ahead with frameworks and architectures for supporting the development of LLM-enabled MAS. The LangChain framework has supplied considerable support for multi-agent LLM programming \cite{topsakal2023creating}. This includes the LangChain Expression language \cite{Darmon2024LCEL}, a domain-specific language that is parsed by the LangChain framework to arrange "chains" of agents to feed inputs and outputs between each other, mostly with support for sequential arrangements. The LangChain expression language is extended by the LangGraph project  \cite{jeong2024study} to manage and arrange branching networks of multiple agents for MAS development.

Major industry vendors are increasingly aligning with community-driven frameworks, actively integrating multi-agent approaches into their products. Notable examples include Microsoft’s AutoGen \cite{wu2023autogen} and Copilot \cite{khan2024microsoft}, as well as Amazon’s Bedrock Agents \cite{amazon2025bedrockagents}. As organisation and open-source projects expand, the challenge of interoperability becomes more pronounced. In response, Anthropic introduced\footnote{And since handed over to the Agentic AI Foundation, directly funded by the Linux Foundation} the Model Context Protocol (MCP), an open standard designed to enable LLMs to connect and interact seamlessly with external tools, data sources, and services \cite{hou2025model}.

Most leading cloud compute suppliers and Agentic AI vendors are rushing to develop similar tooling and support within their own proprietary environments, either based off or heavily influenced by the open-standards and initiatives. At the time of writing, Microsoft has it's Agentic AI development tooling in beta phase in Azure AI Foundry \cite{microsoft2025aifoundry}, and OpenAI offers it's custom '\emph{GPT}'s for users to specialise agents that they can then integrate with wider protocols and environments \cite{eloundou2024gpts}.

\subsection{Research Gap} \label{sect:gap}

Our review reveals a disconnect between two dominant strands of academic literature: \textbf{(i)} highly specialised LLM-enabled MAS designs evaluated in narrow contexts, and \textbf{(ii)} visionary frameworks proposing global, decentralised agent ecosystems. In contrast, industry efforts are rapidly converging on practical MAS development, driven by advanced tooling and open standards that enable organisations to address domain-specific challenges at scale whilst mitigating the inherent challenges that arise from MAS.

This divergence highlights the need for formalisation of emerging design patterns that underpin these industry-driven solutions. While such patterns are increasingly adopted in practice, they remain under-explored in academic discourse. In this paper, we address this gap by documenting a common MAS design paradigm and evaluating its utility through empirical tests across three case studies. Our findings provide insights into the strengths and limitations of this approach and identify complexities, such as alignment and scalability, that traditional architectures often avoid.

\section{Motivating the use of LLM enabled MAS design patterns in organisations} \label{sect:keychallenges}

Organisations face a multitude of challenges in managing their data-assets effectively \cite{sarker2022comprehensive, escobar2021quality}. In particular, many have accumulated vast quantities of data \cite{Pallardy2024howmuchdata}, yet struggle to organise, govern and interpret it in a way that generates meaningful insights for decision-making and supports automation \cite{ugarte2021data, Sellbery2025howmuchdata}, with as little as 16\% being classed as data-driven according to Lewkowicz J. \cite{Lewkowicz2024datafocus}. 
In this section we examine three core data challenges confronting organisations that MAS is positioned to address: Disparate Datasets, Unstructured Datasets, and Domain Specific Data Context.

\subsection{Disparate Data}

As organisations grow in scale and complexity, they often encounter significant challenges associated with data fragmentation and siloed information systems \cite{carruthers2022breaking}. For large organisations in sectors such as Finance \cite{Johnny2023datasilosgovernance} and Government \cite{Kelly2025legacygovernment}, data is often generated and stored across multiple departments, locations, and platforms, each of which may employ different data standards, formats, and schemas. Over time, this unmanaged decentralisation leads to inconsistencies and incompatibilities that make data integration, governance, and cross-functional analysis increasingly difficult. Traditional solutions, such as static data pipelines or schema-mapping tools, often struggle to cope with the dynamic and heterogeneous nature of these distributed data environments, especially when applied retrospectively \cite{bakis2007towards}.

LLM-enabled MAS offers a promising alternative to overcoming this challenge. Through the use of specialist AI agents capable of interpreting, aligning, and translating disparate data schemas in real time, MAS can dynamically integrate information from multiple, previously isolated sources. These agents can autonomously identify relationships between datasets, resolve schema mismatches, and harmonise data structures to enable unified access and analysis. This adaptability allows agent-based systems to act as intelligent intermediaries, facilitating improved navigation across diverse, distributed, and evolving data landscapes that would usually consume significant workforce time.

\subsection{Unstructured Data}

When data is structured and well-organised at its source, it can be readily managed, queried, and analysed to generate reliable insights. For example, sensor outputs typically conform to predefined schemas \cite{dunning2015time}, producing consistently formatted records, while transactional data stored in relational databases adheres to established data models and validation rules \cite{date1989guide}. Such structured datasets are well-suited to conventional analytical and automation techniques \cite{IBM2025structured}.

However, organisations are increasingly faced with large volumes of unstructured or semi-structured data, estimated to comprise as much as 90\% of enterprise information \cite{muscolino2023untapped}. These datasets lack consistent schemas or formats \cite{tredinnick2024managing} and include examples such as interview transcripts, emails, reports, and free-form text, all characterised by ambiguity, context dependency, and linguistic variability \cite{inmon2007tapping}. These properties make integration and interpretation significantly more challenging, often requiring advanced natural language processing (NLP) and semantic reasoning techniques to extract meaningful structure \cite{gharehchopogh2011analysis, li2022neural}.

LLM-based agents offer a powerful capability to address these challenges by performing sophisticated NLP and semantic reasoning on unstructured datasets \cite{deng2025unstructured}. When orchestrated within a MAS, these agents can operate seamlessly across both structured and unstructured data sources, creating an integrated Single Information Environment (SIE) through which information can be navigated, queried, and understood in real time \cite{li2024llm, renney2025reimagining}.

\subsection{Domain Specific Problems}

Across industries, organisations often face similar categories of challenges, yet the specific context, constraints, and requirements within each domain introduce nuances that demand tailored solutions \cite{mohagheghi2010evaluating}. For instance, systems built for healthcare \cite{arunprasath2024improving}, defence \cite{yuvathasecure}, or finance \cite{Kazmi2025dataregsfinance} may share a need for secure data retrieval or analysis, but differ substantially in their data semantics, regulatory obligations, and interpretive logic. Developing and maintaining such bespoke systems traditionally requires specialised software engineering expertise and personnel capable of aligning technical solutions with domain-specific needs, creating dependencies that can limit scalability and agility \cite{mernik2005and}.

LLM-enabled MAS provides adaptive AI that when linked to contextual resources and tooling, is quickly configured to meet domain specific needs. By configuring specialist agents with domain-relevant knowledge, ontologies, and reasoning capabilities, a MAS can be rapidly adapted to new contexts without extensive retraining or redevelopment. These agents can interpret domain-specific data \cite{aryal2024leveraging}, apply contextual understanding \cite{du2024survey}, and generate informed outputs, effectively bridging the gap between generic AI capability and domain-specialised application. As a result, MAS can reduce reliance on human intervention for bespoke problem-solving \cite{cambon2023early}, allowing professionals to focus on higher-level analysis while maintaining a flexible, reusable framework for tackling diverse, domain-specific challenges.


\section{An Emerging Paradigm} \label{sect:paradigm}

Building on the key challenges faced by organisations, this section introduces the foundational elements of the emerging MAS paradigm being documented in this paper. We begin with the latest conceptualisation of specialist autonomous agents, commonly referred to as ReAct Agents and progress towards their integration into collaborative networks. These networks form the basis of broader systems designed to address complex, domain-specific problems.

\subsection{ReAct Agents}
\label{sec:react}

In 2022, the concept of an individual agent operating within a CoT reasoning loop and performing decision-making tasks has been formalised in the definition of the ReAct agent \cite{yao2022react}. The ReAct agent involves the following core components:

\begin{itemize}
    \item \textbf{Tools:} Functions that perform specific tasks, such as querying the Google Search API, accessing an SQL database, executing code via a Python Interpreter \cite{python2025interpreter}, or performing calculations.
    \item \textbf{Reasoning Engine:} The large pre-trained model that powers the system. State-of-the-art LLMs are preferred due to their advanced reasoning capabilities.
    \item \textbf{Agent orchestration:} The overarching system that manages interactions between the LLM and its tools.
    \item \textbf{Memory:} Mechanisms for tracking past reasoning and inputs to inform current decisions. Short-term memory is typically maintained within the prompt context, while long-term memory is stored externally (e.g., in a vector database) and retrieved using semantic similarity to provide relevant context for ongoing tasks.
\end{itemize}

\begin{figure}[H]
\centering
\includegraphics[scale=0.8]{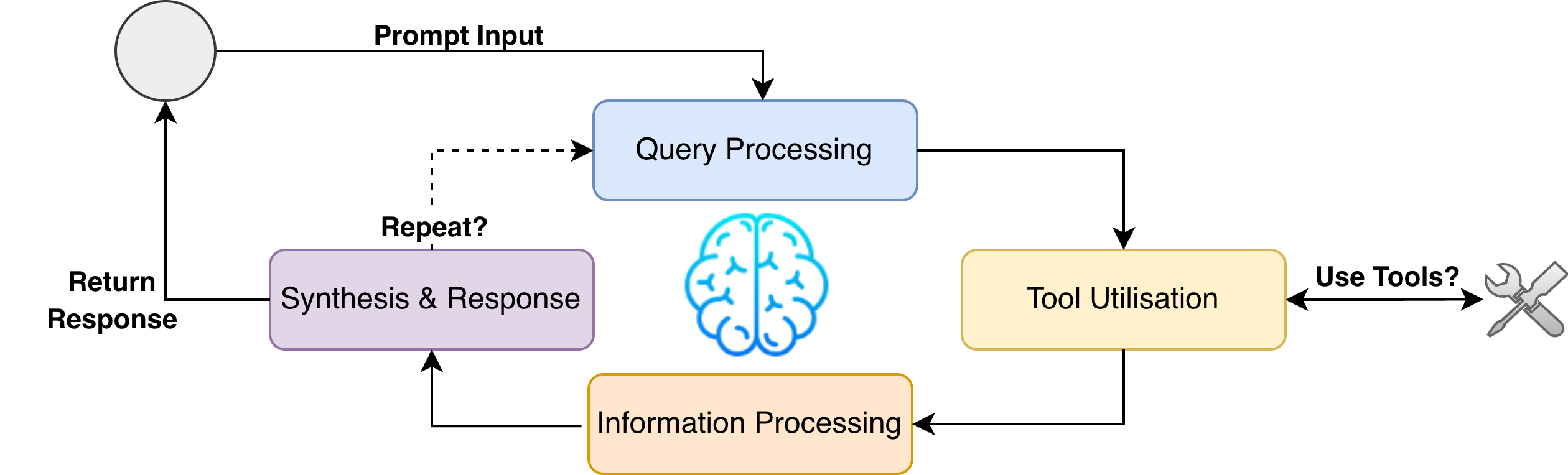}
\caption{Workflow loop of a ReAct agent, depicting iterative reasoning and action execution of key stages, enabling recursive reasoning through chain-of-thought reasoning.}
\label{fig:react-loop}
\end{figure}

Following the workflow loop illustrated in Figure~\ref{fig:react-loop}, such an agent typically operates through the following steps:

\begin{enumerate}
     \item \textbf{Query Processing:} The system processes an input query from another entity (e.g., a user or another agent).
     \item \textbf{Tool Utilization:} The agent selects an appropriate tool (if necessary), prepares its input, executes the tool, and obtains the output.
     \item \textbf{Information Processing:} The agent interprets the tool’s input and output, generates an observation, and determines the next action.
     \item \textbf{Synthesis \& Response:} Using the accumulated information, the agent decides whether it has sufficient context to return a response or whether to repeat the process (returning to Step 1) until a coherent answer is formed, as per CoT reasoning.
\end{enumerate}

With this definition of a specialised ReAct Agent, capable of recursive reasoning and action execution, we can begin to consider how such agents can be organised as part of a broader MAS.

\subsection{Multi-Agent Systems (MAS)} \label{sec:ss1}

While a single agent can be highly effective when tailored to perform a specialised task \cite{ferrag2025llm}, increasing its scope by adding too many available tools often introduces complexity and degrades performance. Empirical evidence shows that agent performance often degrades when the toolkit expands beyond 8–12 tools, due to context-window overload and cognitive interference \cite{hinthorn2025benchmark, prompt2025MCPoverload}. This underpins the rationale for delegating capabilities to specialist agents in a MAS, rather than a single, overburdened general model.

With MAS, capabilities can be distributed across specialist agents that communicate and share reasoning with one another. This approach preserves the effectiveness of individual agents while enabling the system to tackle more complex, multi-faceted problems. The resulting MAS offers several key benefits: \textbf{(i) Modularity:} Independent agents simplify development, testing, and maintenance; \textbf{(ii) Specialisation:} Expert agents can focus on specific domains, improving accuracy and efficiency; \textbf{(iii) Control:} Each agent can be explicitly configured and managed, offering finer grained control to improve alignment with system objectives; \textbf{(iv) Economic:} Tasks are allocated only to relevant agents, meaning agents that are not required remain inactive. This modular activation reduces computational overhead and operational costs. Furthermore, agents can be distilled into smaller, more efficient versions optimised for their specialised tasks \cite{yang2024effective}.

\begin{figure}[H]
\centering
\includegraphics[scale=0.9]{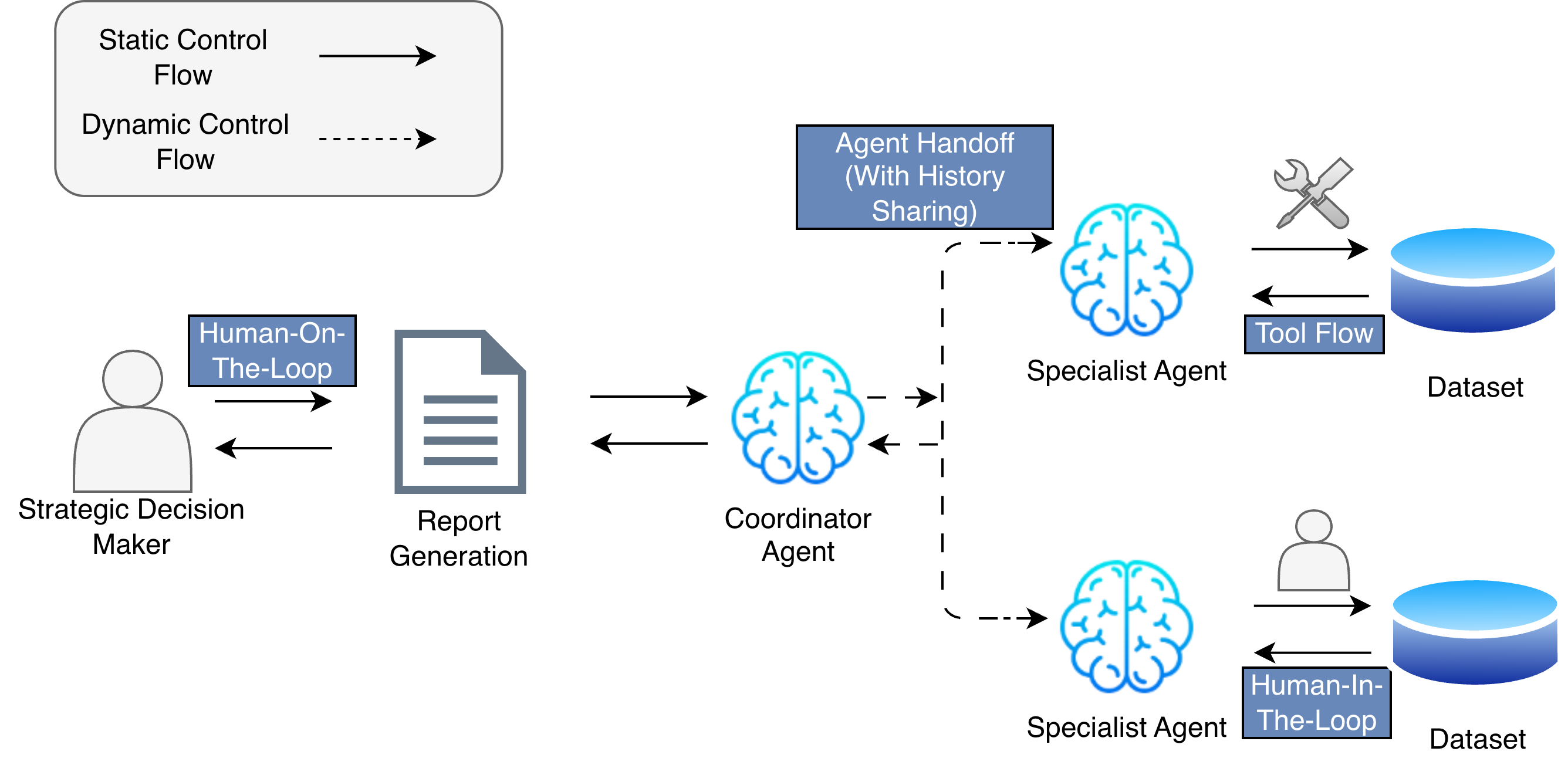}
\caption{Abstract representation of a Multi-Agent System configured as a Single Information Environment (SIE). The diagram illustrates dynamic control flow, agent handoff strategies, and human-on-the-loop oversight for coordinated data retrieval and decision-making.}
\label{fig:sie-arch}
\end{figure}

Defining communication mechanisms is a critical step in assembling a MAS. Building on the concept of specialist ReAct Agents introduced in Section~\ref{sec:react}, Table~\ref{tab:mas} covers key components of MAS architectures that will be referred back to in this paper, including control flow strategies, interaction styles, history-sharing policies, and network configurations to coordinate agent behaviour effectively. Figure~\ref{fig:sie-arch} illustrates these components within an abstract Single Information Environment (SIE) configuration. In this example, a strategic decision-maker interacts with the system as a human-on-the-loop, requesting reports through a coordinator agent. The coordinator employs a dynamic control flow to determine, at runtime, which specialist data-retrieval agents should be engaged via agent handoff with history sharing. One specialist agent uses an explicit tool flow to fetch data, while another operates under an explicit human-in-the-loop control flow for data acquisition. This diagram syntax will be used throughout the remainder of the paper to convey MAS network configurations concisely.

\begin{table}[]
    \centering
    \resizebox{\textwidth}{!}{\begin{tabular}{r|l}
        Aspect & Description\\
        \hline
         Control Flow & \makecell[l]{\textbf{Explicit:} Predefined sequence of agent interactions via graph edges.\\\textbf{Dynamic:} LLM decides flow at runtime.}\\
         \hline
         Interaction Styles & \makecell[l]{\textbf{Handoff:} Passes prompt to a new agent.\\\textbf{Tool Flow:} Leaf nodes call tools; intermediate nodes manage handoffs.}\\
         \hline
         History Sharing & \makecell[l]{\textbf{Full reasoning trace:} Agents share a labelled log of all agent actions,\\communications and reasoning steps as part of their handoff prompts.\\\textbf{Final results only:} Internal steps between input and results are not\\exchanged in handoff between agents.}\\
         \hline
         Network Configurations & \makecell[l]{\textbf{Supervisor:} Leaf node agents coordinated by supervisor agents.\\\textbf{Swarm:} A decentralised structure without any hierarchy where agents\\dynamically hand off tasks to each other based on specialisation.\\\textbf{Hierarchical:} A layered architecture that organises agents under multiple\\supervisory tiers for scalability and structured control.\\\textbf{Single Information Environment (SIE):} A data-centric design where agents\\specialise in unique datasets, with coordinators routing queries to the\\appropriate agent..}\\
         \hline
         Human Interaction & \makecell[l]{\textbf{Human-in-the-loop:} Approve/reject, review tool calls, interrupts.\\\textbf{Human-on-the-loop:} Oversight only.}\\
    \end{tabular}}
    \caption{Stack of common definitions for key components used for defining MAS architectures.}
    \label{tab:mas}
\end{table}

\section{Case-Series} \label{sec:caseseries}

This section applies the emerging paradigm that leverages MAS patterns and LLMs as core reasoning engines for each agent in a series of three proofs-of-concept (PoCs) and pilot studies outlined in Table~\ref{tab:case} addressing real-world challenges across multiple industries. These include telecommunications, government defence, national heritage, and utilities, demonstrating the versatility and scalability of MAS in diverse, high-stakes domains.

\begin{table}[H]
    \centering
    \resizebox{\textwidth}{!}{\begin{tabular}{|c|c|c|c|c||}
        \hline
        \textbf{ID}  & \textbf{Domain} & \textbf{MAS Architecture} & \textbf{Evaluation Metrics} \\
        \hline
        CS1 & Telecommunications Security & SIE & Stakeholder Feedback \\
        \hline
        CS2 & National Heritage Sector & SIE & Stakeholder Feedback \& Open-Ended review per Query \\
        \hline
        CS3 & Utilities Sector & Hierarchical & UAT, Likert ratings, categorisation labelling \\
        \hline
    \end{tabular}}
    \caption{Summary table of the case studies covered within this study.}
    \label{tab:case}
\end{table}

Beyond these examples, the approach is also being extended to intelligent and secure data migration within the Government Defence sector, which will be evaluated as part of future work.

\subsection{Telecoms Security - Security Operations Centre Agents}

In this case study, the security division of a major UK telecommunications provider sought to enhance intelligent tooling and data support within its SOC. The organisation’s existing approach to cyber threat intelligence was heavily human-centric, requiring dozens of analysts to manage vast volumes of data \cite{crowley2019SOC}. This approach is increasingly unsustainable, unable to keep pace with the exponential growth and complexity of the global threat landscape \cite{zidan2024assessing}. For additional context, publicly disclosed vulnerabilities, representing only a fraction of SOC data feeds, have surged from a few thousand in 2016 to over 46,000 annually in 2025\footnote{\url{https://www.cvedetails.com/browse-by-date.php}}. Additionally, from the estimated 400 million terabytes of data generated online each day \cite{Loy2025terabytes}, open-source intelligence (OSINT) must be gathered, filtered and analysed. Managing this manually is far beyond the capacity of traditional human-centric analysis \cite{stolovitch2006handbook}.

Compounding this challenge, the cybersecurity domain suffers from inconsistent threat taxonomies, fragmented data sources, and a pronounced Western-centric bias \cite{cristiano2023cybersecurity}. This convergence of data overload, analytical fragmentation, and intelligence gaps creates a critical bottleneck in cyber defence. Forward-thinking organisations, are exploring how LLMs can power specialist agents for automated data retrieval and insight generation within SOC environments.

\begin{figure}[H]
\centering
\includegraphics[scale=0.8]{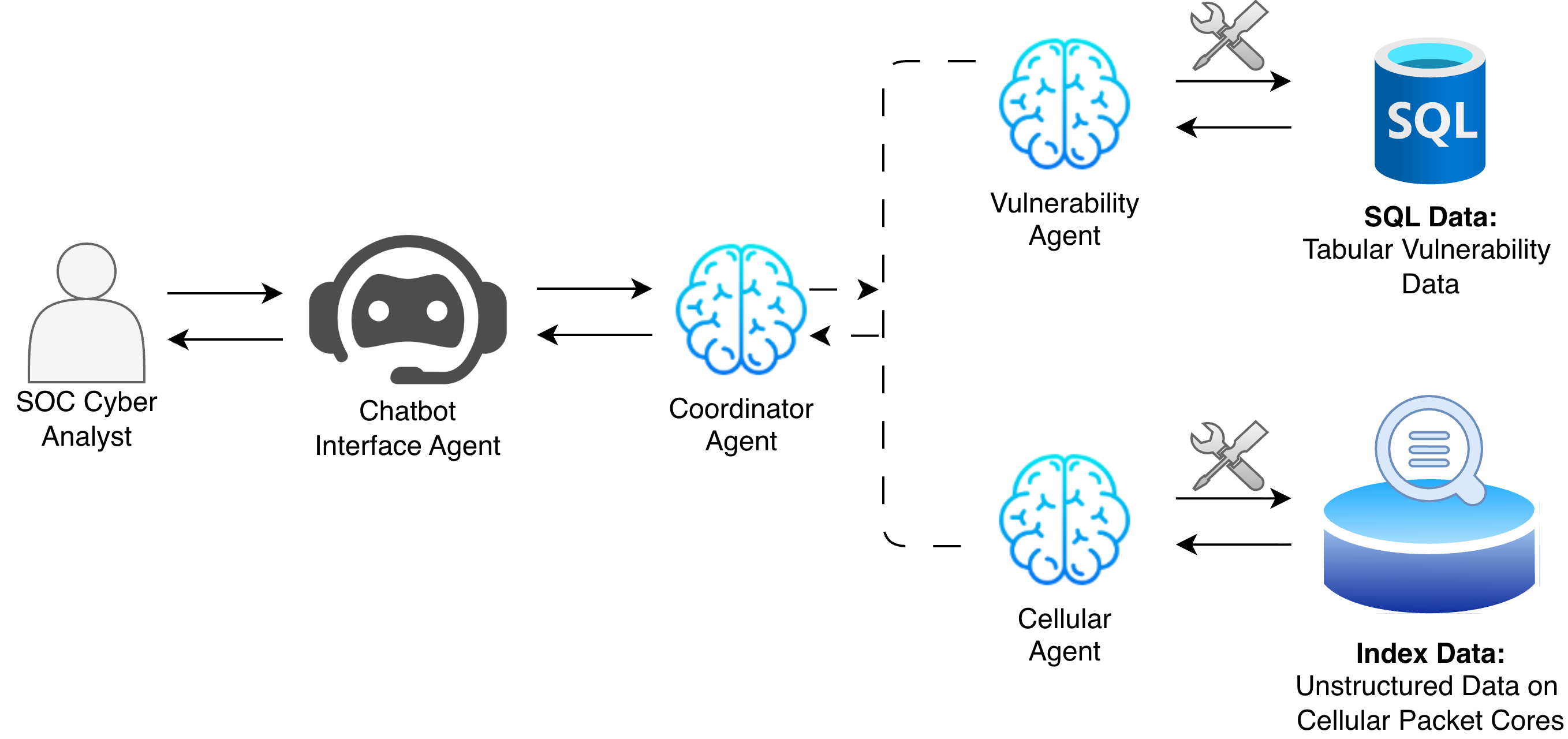}
\caption{MAS architecture deployed in a Telecom SOC case study. The design integrates specialised agents for threat intelligence retrieval and correlation, coordinated through a central supervisor agent, with a conversational interface for analyst interaction.}
\label{fig:CS2-fig}
\end{figure}

This study explored how a MAS architecture could support SOC analysts by automating critical tasks such as data analysis and cross-correlation. The prototype, shown in Figure~\ref{fig:CS2-fig}, implemented a SIE to enable seamless integration of structured SOC datasets with unstructured OSINT feeds. To enhance usability, the solution incorporated a conversational chatbot interface, allowing analysts to query and navigate complex data sources intuitively and in real time.

\begin{figure}
\centering
\includegraphics[scale=0.65]{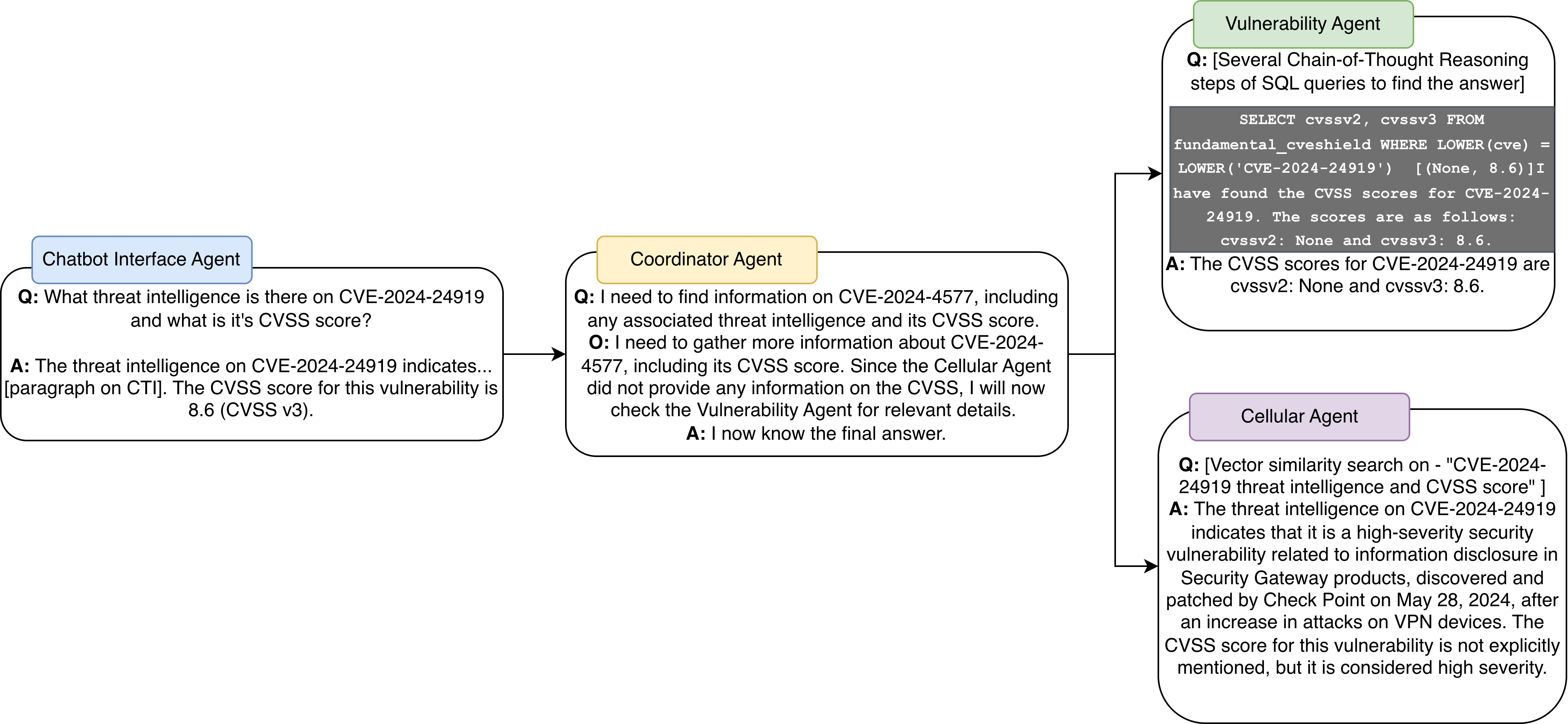}
\caption{Example of MAS reasoning stages in the Telecoms SOC case study. This demonstrates seamless integration of structured and unstructured data sources, supporting efficient and context-rich responses to analyst queries.}
\label{fig:CS2-reasoning}
\end{figure}

Figure~\ref{fig:CS2-reasoning} depicts an example of the reasoning stages logged within the Telecoms SOC case study. In this example, a user initiates a query via the Chatbot Interface Agent, requesting details on a specific CVE and its associated CVSS score. The Coordinator Agent delegates tasks to two specialised agents: the Cellular Agent, which performs vector similarity searches to extract contextual threat intelligence, and the Vulnerability Agent, which executes SQL queries using CoT reasoning to retrieve CVSS scores. This multi-agent workflow demonstrates the MAS paradigm’s ability to integrate heterogeneous data sources, combining structured vulnerability databases with unstructured threat reports, while preserving modularity and adaptability for SOC operations. Furthermore, the Coordinator Agent optimises resource utilisation by recognising when a query requires only a single source and delegating tasks accordingly, thereby reducing unnecessary computational overhead.

To ensure compliance with telecommunications security policies, the MAS solution was deployed as a portable Docker container, enabling secure on-premises implementation. The PoC was tested with key stakeholders and received highly positive feedback on usability and potential impact. Analysts noted that the system could significantly reduce workload by automating data navigation and correlation across heterogeneous sources.

Following the demonstration and subsequent discussion with the Chief Information Security Officer (CISO), they planned further investigation into this approach to augment existing SOC analyst tooling. This feedback indicates the viability of MAS as a method for scaling SOC operations amid increasing cyber complexity. It underscores both the operational benefits for analysts and the strategic potential for senior leadership pursuing digital transformation toward a data-centric SOC model.

\subsection{National Asset Register - Data Sanitation \& Retrieval}

This case study focuses on a large UK national organisation managing extensive assets and inventories across numerous sites. Over time, siloed datasets emerged at every location, each adopting unique data capture methods and structural conventions. This fragmentation resulted in poor data standardisation, making asset inventory management and retrieval highly inconsistent and inaccessible for staff. Consequently, each site’s dataset required a different interpretation approach, complicating integration efforts.

While a long-term initiative to standardise data storage and schema design was planned, an interim solution was needed to enable seamless access and querying of asset information across all UK sites. To address this, a MAS PoC was developed using the design paradigm outlined in this paper. The solution implemented an SIE MAS architecture focused on bridging inconsistencies between structured and unstructured datasets.

\begin{figure}[H]
\centering
 \centering
    \begin{minipage}{0.6\textwidth}
        \centering
\includegraphics[scale=0.6]{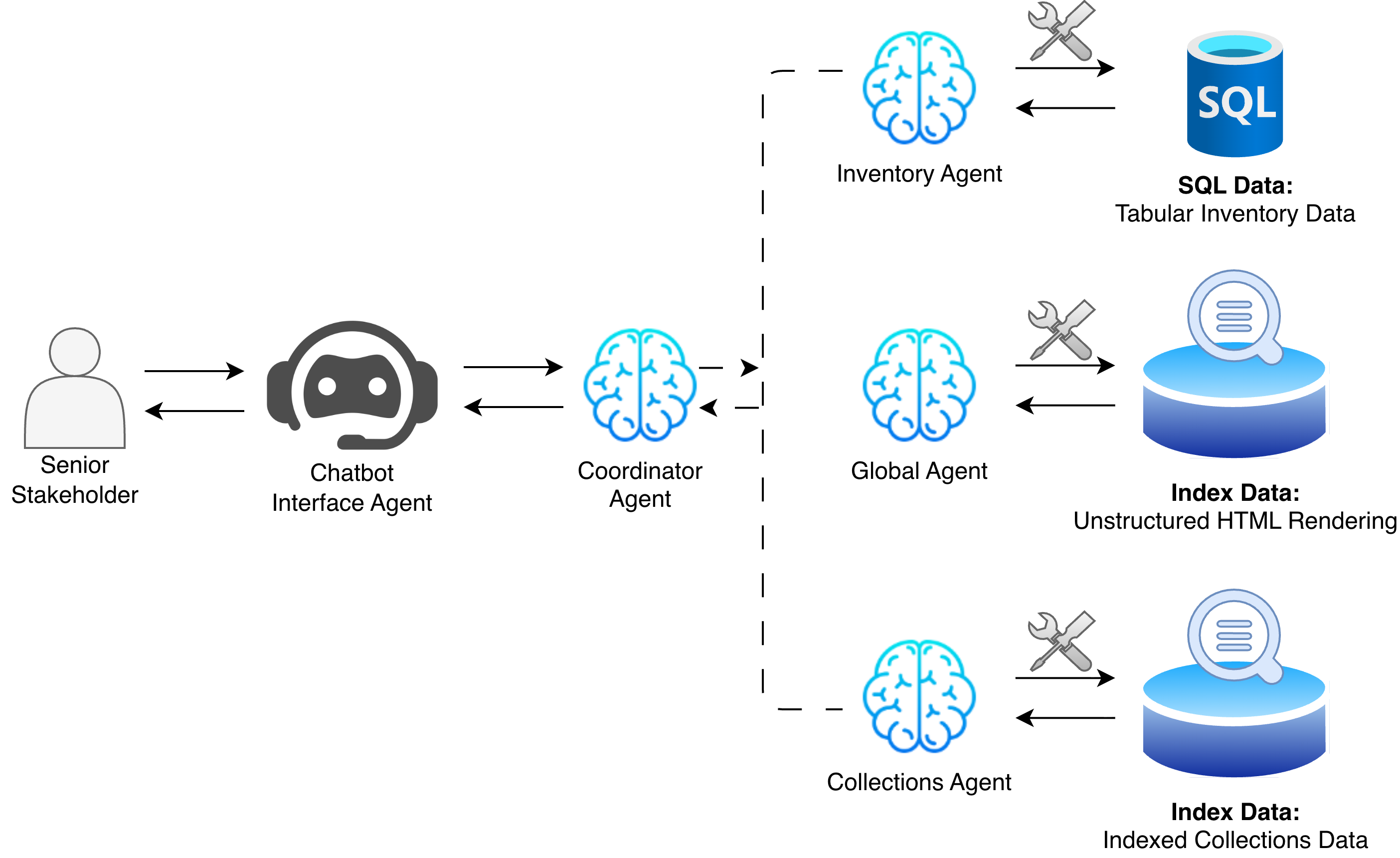} \\
        (\textbf{a})
    \end{minipage}%
    \hfill
    \begin{minipage}{0.35\textwidth}
        \centering
\includegraphics[scale=0.5]{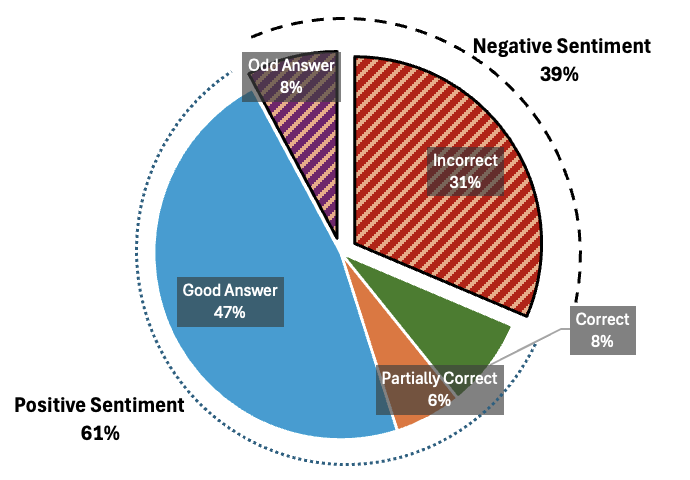} \\
        (\textbf{b})
    \end{minipage}%
    \hfill
\caption{(a) MAS architecture for the National Asset Register case study, featuring dynamic agent orchestration for cross-site data retrieval and sanitation.(b) Sentiment analysis of stakeholder feedback, categorised into positive and negative responses, indicating overall favourable reception but with clear area for improvement.}
\label{fig:case-study-2}
\end{figure}

Through stakeholder discovery sessions and business requirement workshops, the SIE MAS architecture shown in Figure~\ref{fig:case-study-2} \textbf{(a)} was defined. It features a coordinated framework of specialised agents, beginning with a coordinator agent that dynamically hands off prompts to the appropriate data-retrieval specialists that are trained and equipped with data retrieval tooling. To maximise accessibility for non-technical staff, the interface was designed as a natural language chatbot, requiring minimal training and enabling intuitive interaction.

The platform-agnostic design was deployed in the organisation’s Microsoft Azure cloud environment, leveraging native agent services to ensure scalability and compliance. For validation, the PoC was tested by ten senior staff members over one month, with open-ended qualitative feedback collected with a record of comments per use of the solution. Sentiment analysis using Lexicon-Based Matching \cite{inproceedings2019jennifer} categorised responses into positive and negative segments, as shown in Figure~\ref{fig:case-study-2} \textbf{(b)}. Overall, feedback at 61\% was weighted slightly more positively with detail in the feedback indicating how adjustments could be made to quickly improve this. This assumption for the PoC's potential was acknowledged as the stakeholders requested follow-up work to expand the solution. Additionally, planning discussions were held on evolving the internal tool into a customer-facing chatbot to enhance user experience by providing holistic insights into their national assets.

This PoC demonstrated that SIE MAS architectures have potential to unify fragmented datasets without requiring immediate schema standardisation. The conversational interface lowered adoption barriers for non-technical users, while cloud deployment ensured scalability and compliance. However, considering the speed of development taking place within just one month, results indicated that improvements on consistency and reliability are needed. Despite these limitations, strong stakeholder interest in extending the solution to customer-facing applications underscores its potential for broader organisational impact with further testing.

\subsection{Customer Service - Automated Queries \& Complaint Responses}

This case study examines a UK utilities company seeking to improve the efficiency and consistency of its customer service operations, with a particular focus on automating first‑time response (FTR) emails. The organisation had previously commissioned two external contractors to address this challenge; however, neither produced a solution that was either viable or scalable. The present study therefore explored a MAS approach to assess whether targeted automation could deliver measurable performance gains while retaining necessary human oversight.

\begin{figure}[H]
\centering
\includegraphics[scale=0.55]{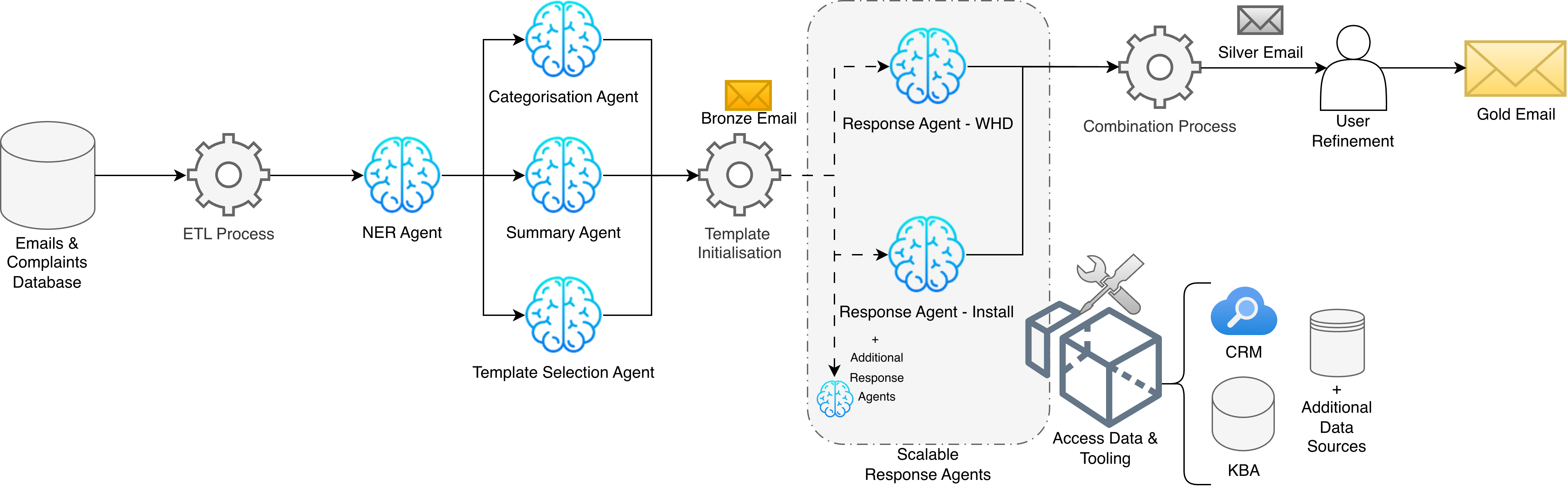}
\caption{End-to-end MAS pipeline for customer service automation, integrating email triage, contextual knowledge retrieval, and response drafting. Human-in-the-loop checkpoints ensure compliance and quality assurance within the automated workflow.}
\label{fig:CS3-arch}
\end{figure}

Following discovery sessions, analysis of the incumbent architecture, and stakeholder consultations, key opportunities for automation were identified in email triage, knowledge retrieval, and response drafting. The original design was adapted as iterative spiral development was conducted \cite{sari2022systematic}. The final resulting MAS architecture (Figure~\ref{fig:CS3-arch}) integrates domain‑specific contextual knowledge from Knowledge Base Articles (KBA) with up‑to‑date customer records from the Customer Relationship Management (CRM) system. Specialised agents perform information extraction from inbound emails, retrieval and synthesis of relevant KBA and CRM context, and generation of draft responses aligned with predefined templates. Human‑in‑the‑loop checkpoints remain embedded for risk‑sensitive decisions and quality assurance, particularly at the user-refinment stage to transition a FTR refined silver email into a gold email ready to send. This configuration constitutes a domain‑specific automation strategy that exceeds the flexibility of conventional scripted or purely rule‑based pipelines.

\vspace{-2cm}

The solution was implemented using a platform‑agnostic design and deployed to Microsoft Azure, utilising Azure’s agent services and hosted within the Microsoft Fabric ecosystem. Integration with Power BI provided operational monitoring and business analytics. In this environment, the MAS produced approximately five FTR emails per minute at an estimated operational cost of \textasciitilde£0.05 per email, with horizontal scalability achievable through distributed, parallel execution. By comparison, the fully manual baseline generated about three FTR emails per minute at a substantially higher \textasciitilde£0.33 per‑email according to the business' estimates, indicating potential improvements in both throughput and cost efficiency.

From an engineering perspective, the MAS paradigm accelerated delivery of initial capabilities. As a point of comparison, the lead developer reported on the project a conventional regular‑expression (regex) categorisation baseline required \textasciitilde 4.5 hours to achieve first testable outputs, whereas an equivalent MAS specialist agent produced comparable first results in less than an hour, reflecting gains in development velocity and extensibility.

\begin{figure}[H]
\centering
 \centering
    \begin{minipage}{0.5\textwidth}
        \centering
\includegraphics[scale=0.5]{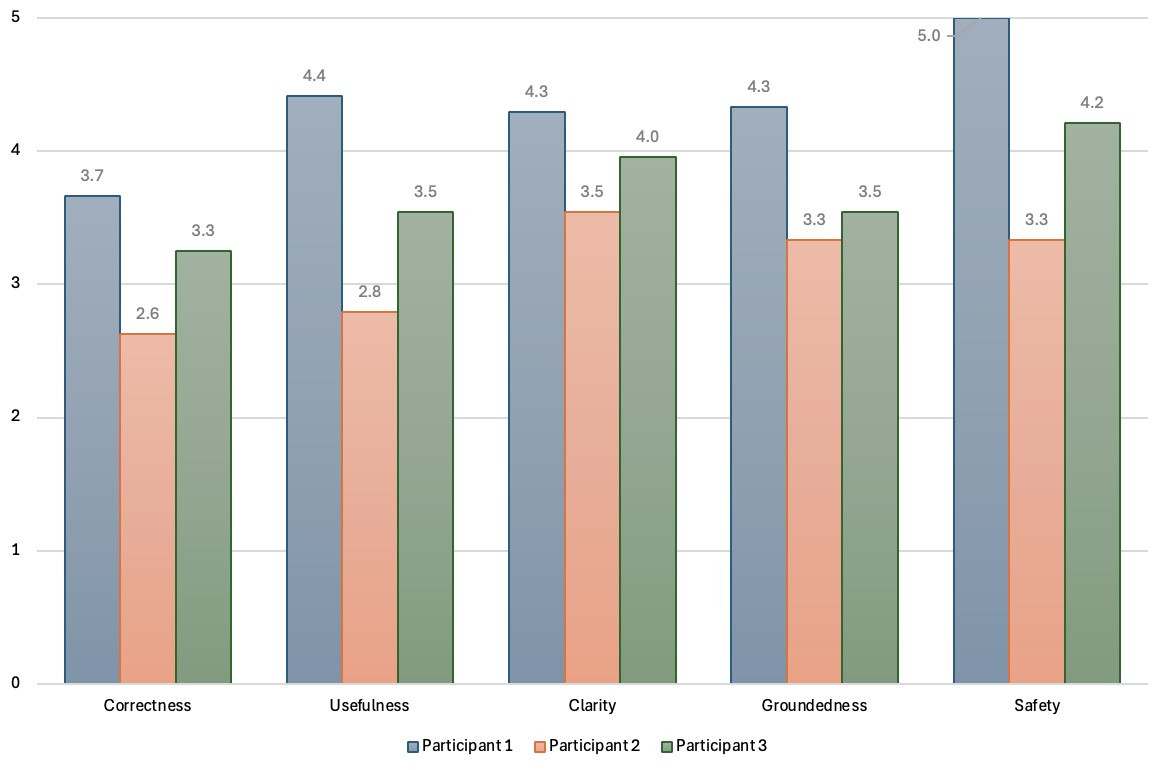} \\
        (\textbf{a})
    \end{minipage}%
    \hfill
    \begin{minipage}{0.35\textwidth}
        \centering
\includegraphics[scale=0.48]{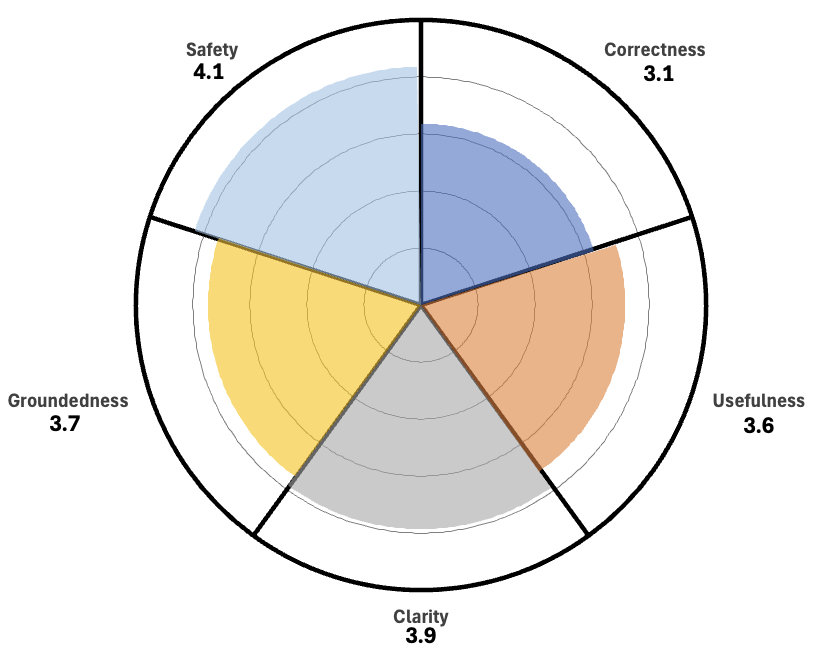} \\
        (\textbf{b})
    \end{minipage}%
    \hfill
\caption{(a) Evaluation results from the UAT per participant, showing Likert-scale ratings across five dimensions: correctness, usefulness, clarity, groundedness, and safety.
(b) Same results aggregated as the mean across participants.}
\label{fig:CS3-results}
\end{figure}

A pilot User Acceptance Test (UAT) involved three participants, who assessed 24 real customer emails. Participants tracked in Figure~\ref{fig:CS3-results} (\textbf{a}) rated system output of FTR using five Likert‑scale dimensions: correctness, usefulness, clarity, groundedness, and safety (1 = poor; 5 = excellent)\footnote{For the interested reader, full details and data are hosted at: \url{https://osf.io/4dzmv/overview}}. The evaluation considered two tasks: (i) categorisation of incoming emails and (ii) FTR generation aligned with organisational templates.

Email categorisation achieved 100\% accuracy when aggregating judgements across all participants. For FTR generation email quality, mean scores (Figure~\ref{fig:CS3-results} (\textbf{b})) on each dimension exceeded 3/5, with clarity and safety averaging \textasciitilde 4/5. Although inter‑participant variability was observed (e.g., safety ratings ranged from 3.3 to 5.0), overall performance surpassed the manual baseline on time‑to‑draft and exhibited promising quality indicators.

This pilot study indicates that an MAS architecture can materially enhance operational efficiency and response consistency in customer service workflows, delivering measurable gains in throughput and unit cost while accelerating development cycles. Notwithstanding the limited scope of evaluation and the identified variability in response quality, stakeholder interest in scaling the system and extending it towards first‑contact resolution underscores its potential as a scalable, adaptable approach to enterprise customer engagement. Future work that prioritises expanded trials is needed, with rigorous A/B testing against manual baselines, and targeted tuning to reduce variance across safety and clarity metrics.

\section{Discussions} \label{sec:discuss}

The case series presented in this paper provides an opportunity to reflect on the practical implications of adopting MAS architectures powered by LLMs as the agent reasoning engine. Drawing on insights from three distinct domains: telecommunications security, national asset management, and customer service automation, the discussion considers the strengths, limitations, and future directions of this emerging paradigm. The following subsections examine three key themes that emerged across all case studies.

\subsection{Rapid Prototype Development}

The case series demonstrates that MAS architectures significantly accelerate prototyping, enabling functional solutions to be developed in weeks rather than months. For instance, in Case Study 3 (Customer Service Automation), two previous contractors failed to deliver any viable solution after several months of development. In contrast, using the MAS approach, the prototype and UAT were completed within one month.

This acceleration is further supported by the platform-agnostic nature of MAS, which allows deployment across on-premises containers and cloud ecosystems without major infrastructure changes, effectively bridging interoperability gaps. However, this speed advantage introduces a critical trade-off: while early-stage development is expedited, transitioning to production requires extensive tuning to address alignment issues and additional compute costs inherent to LLMs, such as variability in outputs and compliance constraints. These challenges may offset initial time savings, underscoring the need for future research to evaluate full lifecycle costs, reliability, and performance benchmarks for MAS at scale.

\subsection{Domain-Specific Adaptability and Efficiency}

A key strength of MAS architectures lies in their adaptability across diverse domains, particularly when combined with LLM reasoning and contextual enrichment through RAG and tool integration. Unlike rigid, rule-based workflows, MAS enables dynamic orchestration of specialised agents, allowing systems to align with domain-specific semantics, compliance requirements, and operational constraints.

This flexibility was evident in the case studies: for example, the SOC prototype automated cross-correlation of heterogeneous threat intelligence sources, a task traditionally requiring extensive manual analysis, while the customer service solution integrated CRM and knowledge base data to generate contextually grounded responses that a specialist would typically have to learn.

Beyond adaptability, MAS delivers efficiency gains by reducing reliance on large development teams and accelerating innovation cycles. These benefits extend beyond speed; they reflect the ability of MAS to scale across problem spaces without wholesale redesign, positioning this paradigm as a reusable foundation for bespoke, high-stakes applications.

The immediate consequence is disruptive organisational impact, as highlighted in emerging enterprise AI research \cite{puavualoaia2023artificial}. Rapid domain-specific adaptability could reshape workforce structures, potentially reducing the need for certain specialised roles and prompting a re-evaluation of organisational strategies.

\subsection{Human Oversight as a Persistent Requirement}

Despite the automation gains observed, all case studies reaffirm the necessity of human involvement, either in-the-loop or on-the-loop, to ensure accountability, compliance, and contextual judgement. As emphasised by Brown \cite{brown2024impactai}, “The future of leadership lies in complementarity [AI], not replacement.” MAS architectures can automate routine tasks and provide actionable insights, but they cannot replicate the nuanced decision-making and ethical considerations inherent to human oversight. In practice, this manifested as supervisory checkpoints in the customer service PoC and strategic decision-maker roles in the SOC and asset management solutions. These findings reinforce the argument that MAS should be positioned as augmentative rather than substitutive technologies within organisational workflows.

\subsection{Key Limitations}

The case series highlights several strengths of the MAS paradigm, such as the initial rapid development lifecycle. However, key limitations are presented through observations by the developers, users and stakeholder feedback. The dependency on LLM reasoning, even when paired with domain specific context via RAG and other tools, inherently has risks of hallucination, grounding errors and and interpretability that can propagate throughout the MAS. This was apparent in the results from Case study 2 and 3, where measurements of reliability and correctness indicate room for improvement. Extended effort is needed to create robust design patterns and guardrails that scale solutions beyond initial rapid development to transition the prototypes into production/release. These issues become more pronounced in high-stakes domains such as cybersecurity or highly regulated industries, reinforcing the need for robust validation pipelines and human-in-the-loop oversight.

Finally, while MAS architectures offer platform-agnostic deployment flexibility, they remain constrained by compute requirements and dependency management. Organisations with limited infrastructure or strict data residency policies may face integration hurdles, particularly when scaling across heterogeneous environments.

\section{Conclusion} \label{sec:conclusion}

This study formalised emerging design patterns for LLM-enabled MAS and evaluated their practical utility across diverse domains through three real-world case studies. The results demonstrate that MAS architectures can significantly accelerate prototyping, reduce development overhead, and deliver adaptable solutions for complex, domain-specific challenges. By leveraging modularity, specialisation, and dynamic orchestration, MAS offers a reusable paradigm that bridges gaps between rigid rule-based workflows and bespoke engineering approaches.

The findings also reinforce persistent limitations. Variability in LLM behaviour, risks of hallucination, and interpretability challenges remain critical barriers to production maturity, particularly in high-stakes or regulated environments. Furthermore, while MAS architectures exhibit strong potential for scalability and interoperability, they introduce dependencies on compute resources and governance frameworks that organisations must address.

Future research will focus on iterative user studies to rigorously evaluate and refine this approach, informing the development of enhanced design features that meet stringent requirements. These efforts will prioritise strengthening validation pipelines and implementing robust guardrails \cite{jai.2025.069841} to mitigate reliability and safety risks.

\textbf{Availability of Data and Materials:} The data supporting the findings of this study are openly available from the Open Science Framework at \url{https://osf.io/4dzmv/overview}.

\textbf{Ethics Approval:} Ethical approval for this study was obtained from the Kaze Consulting review board, and all research protocols were conducted in accordance with company guidelines. All data collected was anonymised and securely stored in compliance with company policies and data protection regulations to ensure the confidentiality and safety of participant information.

\textbf{Conflicts of Interest:} The authors declare no conflicts of interest to report regarding the present study.

\bibliographystyle{unsrt}  
\bibliography{references}

@inproceedings{10.5555/1624861.1624983,
author = {Lesser, Victor R. and Corkill, Daniel D.},
title = {The application of artificial intelligence techniques to cooperative distributed processing},
year = {1979},
isbn = {0934613478},
publisher = {Morgan Kaufmann Publishers Inc.},
address = {San Francisco, CA, USA},
booktitle = {Proceedings of the 6th International Joint Conference on Artificial Intelligence - Volume 1},
pages = {537–540},
numpages = {4},
location = {Tokyo, Japan},
series = {IJCAI'79}
}

@article{hewitt1977viewing,
  title={Viewing control structures as patterns of passing messages},
  author={Hewitt, Carl},
  journal={Artificial intelligence},
  volume={8},
  number={3},
  pages={323--364},
  year={1977},
  publisher={Elsevier}
}

@inproceedings{renney2025reimagining,
  title={Reimagining the Data Landscape: A Multi-Agent Paradigm for Data Interfacing},
  author={Renney, Harri and Nethercott, Max and Williams, Olivia and Evetts, Jayden and Lang, James},
  booktitle={2025 8th International Conference on Data Science and Machine Learning Applications (CDMA)},
  pages={114--119},
  year={2025},
  organization={IEEE}
}

@article{sastry2024computing,
  title={Computing power and the governance of artificial intelligence},
  author={Sastry, Girish and Heim, Lennart and Belfield, Haydn and Anderljung, Markus and Brundage, Miles and Hazell, Julian and O'keefe, Cullen and Hadfield, Gillian K and Ngo, Richard and Pilz, Konstantin and others},
  journal={arXiv preprint arXiv:2402.08797},
  year={2024}
}

@article{naveed2025comprehensive,
  title={A comprehensive overview of large language models},
  author={Naveed, Humza and Khan, Asad Ullah and Qiu, Shi and Saqib, Muhammad and Anwar, Saeed and Usman, Muhammad and Akhtar, Naveed and Barnes, Nick and Mian, Ajmal},
  journal={ACM Transactions on Intelligent Systems and Technology},
  volume={16},
  number={5},
  pages={1--72},
  year={2025},
  publisher={ACM New York, NY}
}

@article{vaswani2017attention,
  title={Attention is all you need},
  author={Vaswani, Ashish and Shazeer, Noam and Parmar, Niki and Uszkoreit, Jakob and Jones, Llion and Gomez, Aidan N and Kaiser, {\L}ukasz and Polosukhin, Illia},
  journal={Advances in neural information processing systems},
  volume={30},
  year={2017}
}

@article{abbas2015organization,
  title={Organization of multi-agent systems: an overview},
  author={Abbas, Hosny Ahmed and Shaheen, Samir Ibrahim and Amin, Mohammed Hussein},
  journal={arXiv preprint arXiv:1506.09032},
  year={2015}
}

@inproceedings{pan2025multiagent,
  title={Why do multiagent systems fail?},
  author={Pan, Melissa Z and Cemri, Mert and Agrawal, Lakshya A and Yang, Shuyi and Chopra, Bhavya and Tiwari, Rishabh and Keutzer, Kurt and Parameswaran, Aditya and Ramchandran, Kannan and Klein, Dan and others},
  booktitle={ICLR 2025 Workshop on Building Trust in Language Models and Applications},
  year={2025}
}

@article{dorri2018multi,
  title={Multi-agent systems: A survey},
  author={Dorri, Ali and Kanhere, Salil S and Jurdak, Raja},
  journal={Ieee Access},
  volume={6},
  pages={28573--28593},
  year={2018},
  publisher={IEEE}
}

@inproceedings{zhanginterpreting,
  title={Interpreting and Improving Large Language Models in Arithmetic Calculation},
  author={Zhang, Wei and Wan, Chaoqun and Zhang, Yonggang and Cheung, Yiu-ming and Tian, Xinmei and Shen, Xu and Ye, Jieping},
  booktitle={Forty-first International Conference on Machine Learning}
}

@article{gomez2023confederacy,
  title={A confederacy of models: A comprehensive evaluation of LLMs on creative writing},
  author={G{\'o}mez-Rodr{\'\i}guez, Carlos and Williams, Paul},
  journal={arXiv preprint arXiv:2310.08433},
  year={2023}
}

@article{fakhoury2024llm,
  title={LLM-based Test-driven Interactive Code Generation: User Study and Empirical Evaluation},
  author={Fakhoury, Sarah and Naik, Aaditya and Sakkas, Georgios and Chakraborty, Saikat and Lahiri, Shuvendu K},
  journal={arXiv preprint arXiv:2404.10100},
  year={2024}
}

@article{brin2023large,
  title={How large language models perform on the United States medical licensing examination: a systematic review},
  author={Brin, Dana and Sorin, Vera and Konen, Eli and Nadkarni, Girish and Glicksberg, Benjamin S and Klang, Eyal},
  journal={MedRxiv},
  pages={2023--09},
  year={2023},
  publisher={Cold Spring Harbor Laboratory Press}
}

@article{katz2024gpt,
  title={Gpt-4 passes the bar exam},
  author={Katz, Daniel Martin and Bommarito, Michael James and Gao, Shang and Arredondo, Pablo},
  journal={Philosophical Transactions of the Royal Society A},
  volume={382},
  number={2270},
  pages={20230254},
  year={2024},
  publisher={The Royal Society}
}

@article{achiam2023gpt,
  title={Gpt-4 technical report},
  author={Achiam, Josh and Adler, Steven and Agarwal, Sandhini and Ahmad, Lama and Akkaya, Ilge and Aleman, Florencia Leoni and Almeida, Diogo and Altenschmidt, Janko and Altman, Sam and Anadkat, Shyamal and others},
  journal={arXiv preprint arXiv:2303.08774},
  year={2023}
}

@article{chowdhery2023palm,
  title={Palm: Scaling language modeling with pathways},
  author={Chowdhery, Aakanksha and Narang, Sharan and Devlin, Jacob and Bosma, Maarten and Mishra, Gaurav and Roberts, Adam and Barham, Paul and Chung, Hyung Won and Sutton, Charles and Gehrmann, Sebastian and others},
  journal={Journal of Machine Learning Research},
  volume={24},
  number={240},
  pages={1--113},
  year={2023}
}

@article{rezaee2015average,
  title={Average consensus over high-order multiagent systems},
  author={Rezaee, Hamed and Abdollahi, Farzaneh},
  journal={IEEE Transactions on Automatic Control},
  volume={60},
  number={11},
  pages={3047--3052},
  year={2015},
  publisher={IEEE}
}

@article{sarker2022comprehensive,
  title={A comprehensive review on big data for industries: challenges and opportunities},
  author={Sarker, Supriya and Arefin, Mohammad Shamsul and Kowsher, Md and Bhuiyan, Touhid and Dhar, Pranab Kumar and Kwon, Oh-Jin},
  journal={IEEE Access},
  volume={11},
  pages={744--769},
  year={2022},
  publisher={IEEE}
}

@article{escobar2021quality,
  title={Quality 4.0: a review of big data challenges in manufacturing},
  author={Escobar, Carlos A and McGovern, Megan E and Morales-Menendez, Ruben},
  journal={Journal of Intelligent Manufacturing},
  volume={32},
  number={8},
  pages={2319--2334},
  year={2021},
  publisher={Springer}
}

@misc{Pallardy2024howmuchdata,
  author = {Carrie Pallardy},
  title = {How Much Data Is Too Much for Organizations to Derive Value?},
  year = {2024},
  note = {Last accessed 13 November 2025},
  url = {https://www.informationweek.com/data-management/how-much-data-is-too-much-for-organizations-to-derive-value-}
}

@book{ugarte2021data,
  title={The Data Mirage: Why Companies Fail to Actually Use Their Data},
  author={Ugarte, Ruben},
  year={2021},
  publisher={Business Expert Press}
}

@misc{Sellbery2025howmuchdata,
  author = {Sellbery},
  title = {Why Most Companies Are Still Struggling to Use Their Data Effectively},
  year = {2025},
  note = {Last accessed 13 November 2025},
  url = {https://sellbery.com/blog/why-most-companies-are-still-struggling-to-use-their-data-effectively/}
}

@misc{Lewkowicz2024datafocus,
  author = {Jakub Lewkowicz},
  title = {While most companies focus on data, only about 16\% are ‘data-driven’},
  year = {2024},
  note = {Last accessed 13 November 2025},
  url = {https://sdtimes.com/data/while-most-companies-focus-on-data-only-about-16-are-data-driven/}
}

@misc{Kiela2024RAG,
  author = {Douwe Kiela},
  title = {RAG Agents in Prod: 10 Lessons We Learned — Douwe Kiela, creator of RAG},
  year = {2025},
  note = {Last accessed 13 November 2025},
  url = {https://www.youtube.com/watch?v=kPL-6-9MVyA&t=267s}
}

@article{wei2022chain,
  title={Chain-of-thought prompting elicits reasoning in large language models},
  author={Wei, Jason and Wang, Xuezhi and Schuurmans, Dale and Bosma, Maarten and Xia, Fei and Chi, Ed and Le, Quoc V and Zhou, Denny and others},
  journal={Advances in neural information processing systems},
  volume={35},
  pages={24824--24837},
  year={2022}
}

@article{singh2024rethinking,
  title={Rethinking interpretability in the era of large language models},
  author={Singh, Chandan and Inala, Jeevana Priya and Galley, Michel and Caruana, Rich and Gao, Jianfeng},
  journal={arXiv preprint arXiv:2402.01761},
  year={2024}
}

@article{mesko2023imperative,
  title={The imperative for regulatory oversight of large language models (or generative AI) in healthcare},
  author={Mesk{\'o}, Bertalan and Topol, Eric J},
  journal={NPJ digital medicine},
  volume={6},
  number={1},
  pages={120},
  year={2023},
  publisher={Nature Publishing Group UK London}
}

@article{lewis2020retrieval,
  title={Retrieval-augmented generation for knowledge-intensive nlp tasks},
  author={Lewis, Patrick and Perez, Ethan and Piktus, Aleksandra and Petroni, Fabio and Karpukhin, Vladimir and Goyal, Naman and K{\"u}ttler, Heinrich and Lewis, Mike and Yih, Wen-tau and Rockt{\"a}schel, Tim and others},
  journal={Advances in Neural Information Processing Systems},
  volume={33},
  pages={9459--9474},
  year={2020}
}

@inproceedings{topsakal2023creating,
  title={Creating large language model applications utilizing langchain: A primer on developing llm apps fast},
  author={Topsakal, Oguzhan and Akinci, Tahir Cetin},
  booktitle={International Conference on Applied Engineering and Natural Sciences},
  volume={1},
  number={1},
  pages={1050--1056},
  year={2023}
}

@Misc{Darmon2024LCEL,
author = {Tom Darmon},
title = {Unleashing the Power of LangChain Expression Language (LCEL): from proof of concept to production},
howpublished = {\textsc{url:}~\url{https://www.artefact.com/blog/unleashing-the-power-of-langchain-expression-language-lcel-from-proof-of-concept-to-production/}},
month = {1},
year = {2024}
}

@article{jeong2024study,
  title={A Study on the Implementation Method of an Agent-Based Advanced RAG System Using Graph},
  author={Jeong, Cheonsu},
  journal={arXiv preprint arXiv:2407.19994},
  year={2024}
}

@article{wu2023autogen,
  title={Autogen: Enabling next-gen llm applications via multi-agent conversation framework},
  author={Wu, Qingyun and Bansal, Gagan and Zhang, Jieyu and Wu, Yiran and Zhang, Shaokun and Zhu, Erkang and Li, Beibin and Jiang, Li and Zhang, Xiaoyun and Wang, Chi},
  journal={arXiv preprint arXiv:2308.08155},
  year={2023}
}

@incollection{khan2024microsoft,
  title={Microsoft Copilot Studio},
  author={Khan, Adeel},
  booktitle={Introducing Microsoft Copilot for Managers: Enhance Your Team's Productivity and Creativity with Generative AI-Powered Assistant},
  pages={621--694},
  year={2024},
  publisher={Springer}
}

@article{yao2024tree,
  title={Tree of thoughts: Deliberate problem solving with large language models},
  author={Yao, Shunyu and Yu, Dian and Zhao, Jeffrey and Shafran, Izhak and Griffiths, Tom and Cao, Yuan and Narasimhan, Karthik},
  journal={Advances in Neural Information Processing Systems},
  volume={36},
  year={2024}
}

@article{yao2022react,
  title={ReAct: Synergizing Reasoning and Acting in Language Models},
  author={Yao, Shunyu and Zhao, Jeffrey and Yu, Dian and Du, Nan and Shafran, Izhak and Narasimhan, Karthik and Cao, Yuan},
  journal={arXiv preprint arXiv:2210.03629},
  year={2022}
}

@article{lin2024llm,
  title={When llm-based code generation meets the software development process},
  author={Lin, Feng and Kim, Dong Jae and others},
  journal={arXiv preprint arXiv:2403.15852},
  year={2024}
}

@Misc{kearney2024bots,
author = {{Michael Kearney}},
title = {Bots of the SOC},
howpublished = {\textsc{url:}~\url{https://www.youtube.com/watch?v=WLP5eEqmbxQ}},
month = {8},
year = {2024}
}

@Misc{splunkgh2023BotS,
author = {{Splunk GitHub}},
title = {botsv3},
howpublished = {\textsc{url:}~\url{https://github.com/splunk/botsv3}},
month = {2},
year = {2020}
}

@article{oche2025systematic,
  title={A systematic review of key retrieval-augmented generation (rag) systems: Progress, gaps, and future directions},
  author={Oche, Agada Joseph and Folashade, Ademola Glory and Ghosal, Tirthankar and Biswas, Arpan},
  journal={arXiv preprint arXiv:2507.18910},
  year={2025}
}

@Misc{MIT2025NANDA,
author = {{MIT}},
title = {NANDA: The Internet of AI Agents},
howpublished = {\textsc{url:}~\url{https://nanda.media.mit.edu}},
urldate = {2025-11-22}
}

@Misc{microsoft2025aifoundry,
author = {{Microsoft}},
title = {The AI app and agent factory},
howpublished = {\textsc{url:}~\url{https://ai.azure.com}},
urldate = {2025-11-22}
}

@article{raskar2025beyond,
  title={Beyond dns: Unlocking the internet of ai agents via the nanda index and verified agentfacts},
  author={Raskar, Ramesh and Chari, Pradyumna and Zinky, John and Lambe, Mahesh and Grogan, Jared James and Wang, Sichao and Ranjan, Rajesh and Singhal, Rekha and Gupta, Shailja and Lincourt, Robert and others},
  journal={arXiv preprint arXiv:2507.14263},
  year={2025}
}

@article{eloundou2024gpts,
  title={GPTs are GPTs: Labor market impact potential of LLMs},
  author={Eloundou, Tyna and Manning, Sam and Mishkin, Pamela and Rock, Daniel},
  journal={Science},
  volume={384},
  number={6702},
  pages={1306--1308},
  year={2024},
  publisher={American Association for the Advancement of Science}
}

@inproceedings{10.1609/aaai.v38i21.30361,
author = {Xu, Zhenyu and Sheng, Victor S.},
title = {Detecting AI-generated code assignments using perplexity of large language models},
year = {2024},
isbn = {978-1-57735-887-9},
publisher = {AAAI Press},
url = {https://doi.org/10.1609/aaai.v38i21.30361},
doi = {10.1609/aaai.v38i21.30361},
booktitle = {Proceedings of the Thirty-Eighth AAAI Conference on Artificial Intelligence and Thirty-Sixth Conference on Innovative Applications of Artificial Intelligence and Fourteenth Symposium on Educational Advances in Artificial Intelligence},
articleno = {2638},
numpages = {8},
series = {AAAI'24/IAAI'24/EAAI'24}
}

@article{wu2025survey,
  title={A survey on llm-generated text detection: Necessity, methods, and future directions},
  author={Wu, Junchao and Yang, Shu and Zhan, Runzhe and Yuan, Yulin and Chao, Lidia Sam and Wong, Derek Fai},
  journal={Computational Linguistics},
  volume={51},
  number={1},
  pages={275--338},
  year={2025},
  publisher={MIT Press 255 Main Street, 9th Floor, Cambridge, Massachusetts 02142, USA~…}
}

@inproceedings{inproceedings2019jennifer,
author = {Contreras, Jennifer and Ballera, Melvin and Lagman, Ace and Raviz, Jennalyn},
year = {2019},
month = {03},
pages = {},
title = {Lexicon-based Sentiment Analysis with Pattern Matching Application using Regular Expression in Automata},
doi = {10.1145/3301551.3301596}
}

@article{puavualoaia2023artificial,
  title={Artificial intelligence as a disruptive technology—a systematic literature review},
  author={P{\u{a}}v{\u{a}}loaia, Vasile-Daniel and Necula, Sabina-Cristiana},
  journal={Electronics},
  volume={12},
  number={5},
  pages={1102},
  year={2023},
  publisher={MDPI}
}

@article{cmc.2025.064747,
  title={Upholding Academic Integrity amidst Advanced Language Models: Evaluating BiLSTM Networks with GloVe Embeddings for Detecting AI-Generated Scientific Abstracts.},
  author={Popescu-Apreutesei, Lilia-Eliana and Iosupescu, Mihai-Sorin and Necula, Sabina Cristiana and P{\u{a}}v{\u{a}}loaia, Vasile-Daniel},
  journal={Computers, Materials \& Continua},
  volume={84},
  number={2},
  year={2025}
}

@Article{jai.2024.058649,
AUTHOR = {Afam Okonkwo, Pius Onobhayedo, Charles Igah},
TITLE = {Using Artificial Intelligence Techniques in the Requirement Engineering Stage of Traditional SDLC Process},
JOURNAL = {Journal on Artificial Intelligence},
VOLUME = {6},
YEAR = {2024},
NUMBER = {1},
PAGES = {379--401},
URL = {http://www.techscience.com/jai/v6n1/59162},
ISSN = {2579-003X},
DOI = {10.32604/jai.2024.058649}
}

@Article{jai.2025.069841,
AUTHOR = {Eleena Sarah Mathew},
TITLE = {Enhancing Security in Large Language Models: A Comprehensive Review of Prompt Injection Attacks and Defenses},
JOURNAL = {Journal on Artificial Intelligence},
VOLUME = {7},
YEAR = {2025},
NUMBER = {1},
PAGES = {347--363},
URL = {http://www.techscience.com/jai/v7n1/64040},
ISSN = {2579-003X},
DOI = {10.32604/jai.2025.069841}
}

@Article{jai.2025.069226,
AUTHOR = {Charles Munyao, John Ndia},
TITLE = {Natural Language Processing with Transformer-Based Models: A Meta-Analysis},
JOURNAL = {Journal on Artificial Intelligence},
VOLUME = {7},
YEAR = {2025},
NUMBER = {1},
PAGES = {329--346},
URL = {http://www.techscience.com/jai/v7n1/63779},
ISSN = {2579-003X},
DOI = {10.32604/jai.2025.069226}
}

@book{alto2024building,
  title={Building LLM Powered Applications: Create intelligent apps and agents with large language models},
  author={Alto, Valentina},
  year={2024},
  publisher={Packt Publishing Ltd}
}

@article{wang2024beyond,
  title={Beyond direct diagnosis: LLM-based multi-specialist agent consultation for automatic diagnosis},
  author={Wang, Haochun and Zhao, Sendong and Qiang, Zewen and Xi, Nuwa and Qin, Bing and Liu, Ting},
  journal={arXiv preprint arXiv:2401.16107},
  year={2024}
}

@article{kalyuzhnaya2025llm,
  title={LLM Agents for Smart City Management: Enhancing Decision Support Through Multi-Agent AI Systems.},
  author={Kalyuzhnaya, Anna and Mityagin, Sergey and Lutsenko, Elizaveta and Getmanov, Andrey and Aksenkin, Yaroslav and Fatkhiev, Kamil and Fedorin, Kirill and Nikitin, Nikolay O and Chichkova, Natalia and Vorona, Vladimir and others},
  journal={Smart Cities (2624-6511)},
  volume={8},
  number={1},
  year={2025}
}

@article{zhang2025grid,
  title={Grid-agent: An llm-powered multi-agent system for power grid control},
  author={Zhang, Yan and Saber, Ahmad Mohammad and Youssef, Amr and Kundur, Deepa},
  journal={arXiv preprint arXiv:2508.05702},
  year={2025}
}

@article{berti2025emergent,
  title={Emergent abilities in large language models: A survey},
  author={Berti, Leonardo and Giorgi, Flavio and Kasneci, Gjergji},
  journal={arXiv preprint arXiv:2503.05788},
  year={2025}
}

@Misc{amazon2025bedrockagents,
author = {{Amazon}},
title = {Amazon Bedrock Agents},
howpublished = {\textsc{url:}~\url{https://aws.amazon.com/bedrock/agents}},
urldate = {2025-12-10}
}

@article{hou2025model,
  title={Model context protocol (mcp): Landscape, security threats, and future research directions},
  author={Hou, Xinyi and Zhao, Yanjie and Wang, Shenao and Wang, Haoyu},
  journal={arXiv preprint arXiv:2503.23278},
  year={2025}
}

@article{Johnny2023datasilosgovernance,
author = {Johnny, Ricky},
year = {2023},
month = {09},
pages = {},
title = {Addressing Data Silos with Governance Frameworks in Financial Organizations}
}

@Misc{Kelly2025legacygovernment,
author = {{Piers Kelly}},
title = {Legacy Systems in Government: How to Break Data Silos and Innovate},
howpublished = {\textsc{url:}~\url{https://blog.govnet.co.uk/technology/legacy-systems-in-government-how-to-break-data-silos-and-innovate}},
urldate = {2025-12-10},
year={2025}
}

@incollection{carruthers2022breaking,
  title={Breaking Data Silos},
  author={Carruthers, Andrew},
  booktitle={Building the Snowflake Data Cloud: Monetizing and Democratizing Your Data},
  pages={29--50},
  year={2022},
  publisher={Springer}
}

@article{bakis2007towards,
  title={Towards distributed product data sharing environments—Progress so far and future challenges},
  author={Bakis, Nick and Aouad, Ghassan and Kagioglou, Mike},
  journal={Automation in Construction},
  volume={16},
  number={5},
  pages={586--595},
  year={2007},
  publisher={Elsevier}
}

@article{dunning2015time,
  title={Time series databases},
  author={Dunning, Ted and Friedman, Ellen},
  journal={New Ways to Store and Access data},
  year={2015}
}

@book{date1989guide,
  title={A Guide to the SQL Standard},
  author={Date, Chris J},
  year={1989},
  publisher={Addison-Wesley Longman Publishing Co., Inc.}
}

@Misc{IBM2025structured,
author = {{IBM}},
title = {What are the key differences between structured and unstructured data?},
howpublished = {\textsc{url:}~\url{https://www.ibm.com/think/topics/structured-vs-unstructured-data}},
urldate = {2025-12-14},
year={2025}
}

@misc{muscolino2023untapped,
  title={Untapped value: what every executive needs to know about unstructured data},
  author={Muscolino, H and Machado, A and Rydning, J and Vesset, D},
  year={2023},
  publisher={Needham, MA: IDC Research Ltd}
}

@misc{tredinnick2024managing,
  title={Managing unstructured information},
  author={Tredinnick, Luke and Laybats, Claire},
  journal={Business Information Review},
  volume={41},
  number={3},
  pages={90--93},
  year={2024},
  publisher={SAGE Publications Sage UK: London, England}
}

@book{inmon2007tapping,
  title={Tapping into unstructured data: integrating unstructured data and textual analytics into business intelligence},
  author={Inmon, William H and Nesavich, Anthony},
  year={2007},
  publisher={Pearson Education}
}

@inproceedings{gharehchopogh2011analysis,
  title={Analysis and evaluation of unstructured data: text mining versus natural language processing},
  author={Gharehchopogh, Farhad Soleimanian and Khalifelu, Zeinab Abbasi},
  booktitle={2011 5th International Conference on Application of Information and Communication Technologies (AICT)},
  pages={1--4},
  year={2011},
  organization={IEEE}
}

@article{li2022neural,
  title={Neural natural language processing for unstructured data in electronic health records: a review},
  author={Li, Irene and Pan, Jessica and Goldwasser, Jeremy and Verma, Neha and Wong, Wai Pan and Nuzumlal{\i}, Muhammed Yavuz and Rosand, Benjamin and Li, Yixin and Zhang, Matthew and Chang, David and others},
  journal={Computer Science Review},
  volume={46},
  pages={100511},
  year={2022},
  publisher={Elsevier}
}

@article{deng2025unstructured,
  title={Unstructured Data Analysis using LLMs: A Comprehensive Benchmark},
  author={Deng, Qiyan and Li, Jianhui and Chai, Chengliang and Liu, Jinqi and She, Junzhi and Jin, Kaisen and Sun, Zhaoze and Deng, Yuhao and Yuan, Jia and Yuan, Ye and others},
  journal={arXiv preprint arXiv:2510.27119},
  year={2025}
}

@inproceedings{li2024llm,
  title={LLM for Uniform Information Extraction Using Multi-task Learning Optimization},
  author={Li, Ying and Tan, Zhen and Xiao, Weidong},
  booktitle={Asia-Pacific Web (APWeb) and Web-Age Information Management (WAIM) Joint International Conference on Web and Big Data},
  pages={17--29},
  year={2024},
  organization={Springer}
}

@inproceedings{mohagheghi2010evaluating,
  title={Evaluating domain-specific modelling solutions},
  author={Mohagheghi, Parastoo and Haugen, {\O}ystein},
  booktitle={International Conference on Conceptual Modeling},
  pages={212--221},
  year={2010},
  organization={Springer}
}

@article{arunprasath2024improving,
  title={Improving patient centric data retrieval and cyber security in healthcare: privacy preserving solutions for a secure future},
  author={Arunprasath, S and Annamalai, Suresh},
  journal={Multimedia Tools and Applications},
  volume={83},
  number={27},
  pages={70289--70319},
  year={2024},
  publisher={Springer}
}

@article{yuvathasecure,
  title={Secure Data Retrieval \& Recovery System Decentralized Interruption Tolerant Defence Data Network},
  author={Yuvatha, Rathnam and Narayana, J Laxmi and Gouthami, V}
}

@Misc{Kazmi2025dataregsfinance,
author = {{Syed Tatheer Kazmi, Rohma Fatima Qayyum}},
title = {Data Regulations in the UK’s Financial Sector},
howpublished = {\textsc{url:}~\url{https://securiti.ai/data-regulations-in-the-uk-financial-sector/}},
urldate = {2025-12-15},
year={2025}
}

@article{mernik2005and,
  title={When and how to develop domain-specific languages},
  author={Mernik, Marjan and Heering, Jan and Sloane, Anthony M},
  journal={ACM computing surveys (CSUR)},
  volume={37},
  number={4},
  pages={316--344},
  year={2005},
  publisher={ACM New York, NY, USA}
}

@article{aryal2024leveraging,
  title={Leveraging multi-AI agents for cross-domain knowledge discovery},
  author={Aryal, Shiva and Do, Tuyen and Heyojoo, Bisesh and Chataut, Sandeep and Gurung, Bichar Dip Shrestha and Gadhamshetty, Venkataramana and Gnimpieba, Etienne},
  journal={arXiv preprint arXiv:2404.08511},
  year={2024}
}

@article{du2024survey,
  title={A survey on context-aware multi-agent systems: techniques, challenges and future directions},
  author={Du, Hung and Thudumu, Srikanth and Vasa, Rajesh and Mouzakis, Kon},
  journal={arXiv preprint arXiv:2402.01968},
  year={2024}
}

@article{cambon2023early,
  title={Early LLM-based tools for enterprise information workers likely provide meaningful boosts to productivity},
  author={Cambon, Alexia and Hecht, Brent and Edelman, Ben and Ngwe, Donald and Jaffe, Sonia and Heger, Amy and Vorvoreanu, Mihaela and Peng, Sida and Hofman, Jake and Farach, Alex and others},
  journal={Microsoft Research. MSR-TR-2023},
  volume={43},
  year={2023}
}

@Misc{python2025interpreter,
author = {{Python Software Foundation.}},
title = {Using the Python Interpreter},
howpublished = {\textsc{url:}~\url{https://docs.python.org/3/tutorial/interpreter.html}},
urldate = {2025-12-15},
year={2025}
}

@Misc{hinthorn2025benchmark,
author = {{Will Fu-Hinthorn}},
title = {Benchmarking Multi-Agent Architectures},
howpublished = {\textsc{url:}~\url{https://blog.langchain.com/benchmarking-multi-agent-architectures/}},
urldate = {2025-12-15},
year={2025}
}

@Misc{prompt2025MCPoverload,
author = {{PromptForward Team}},
title = {MCP Overload: Why Your LLM Agent Doesn’t Need 20 Tools},
howpublished = {\textsc{url:}~\url{https://promptforward.dev/blog/mcp-overload}},
urldate = {2025-12-15},
year={2025}
}

@article{ferrag2025llm,
  title={From llm reasoning to autonomous ai agents: A comprehensive review},
  author={Ferrag, Mohamed Amine and Tihanyi, Norbert and Debbah, Merouane},
  journal={arXiv preprint arXiv:2504.19678},
  year={2025}
}

@unknown{luyen2025howwell,
author = {Luyen, Le and Abel, Marie-Hélène},
year = {2025},
month = {07},
pages = {},
title = {How Well Do LLMs Predict Prerequisite Skills? Zero-Shot Comparison to Expert-Defined Concepts},
doi = {10.48550/arXiv.2507.18479}
}

@Misc{brown2024impactai,
author = {{Professor Alan Brown}},
title = {The Impact of AI on the Past, Present, and Future of Leadership - Part 1},
howpublished = {\textsc{url:}~\url{https://www.oxford-group.com/insights/the-impact-of-ai-on-leadership/}},
urldate = {2025-12-16},
year={2025}
}

@Misc{Loy2025terabytes,
author = {{Matt Loy}},
title = {How Much Data Is Generated Per Day},
howpublished = {\textsc{url:}~\url{https://www.techbusinessnews.com.au/blog/402-74-million-terrabytes-of-data-is-created-every-day/}},
urldate = {2025-12-16},
year={2025}
}

@book{stolovitch2006handbook,
  title={Handbook of human performance technology: Principles, practices, and potential},
  author={Stolovitch, Harold D and Keeps, Erica J},
  year={2006},
  publisher={John Wiley \& Sons}
}

@Misc{chanen2023learnedAI,
author = {{Ari Chanen}},
title = {What I learned from Bloomberg's experience of building their own LLM},
howpublished = {\textsc{url:}~\url{https://www.linkedin.com/pulse/what-i-learned-from-bloombergs-experience-building-own-chanen-phd/}},
urldate = {2025-12-16},
year={2023}
}

@article{cristiano2023cybersecurity,
  title={Cybersecurity and the politics of knowledge production: towards a reflexive practice},
  author={Cristiano, Fabio and Kurowska, Xymena and Stevens, Tim and Hurel, Louise Marie and Fouad, Noran Shafik and Cavelty, Myriam Dunn and Broeders, Dennis and Liebetrau, Tobias and Shires, James},
  journal={Journal of Cyber Policy},
  volume={8},
  number={3},
  pages={331--364},
  year={2023},
  publisher={Taylor \& Francis}
}

@inproceedings{zidan2024assessing,
  title={Assessing the Challenges Faced by Security Operations Centres (SOC)},
  author={Zidan, Kamal and Alam, Abu and Allison, Jordan and Al-sherbaz, Ali},
  booktitle={Future of Information and Communication Conference},
  pages={256--271},
  year={2024},
  organization={Springer}
}

@Misc{crowley2019SOC,
author = {{Chris Crowley}},
title = {Common and Best Practices for Security Operations Centers: Results of the 2019 SOC Survey},
publisher = {SANS Institute},
urldate = {2025-12-16},
year={2019}
}

@article{sari2022systematic,
  title={The Systematic Literature Review of the spiral development model: Topics, trends, and application areas},
  author={Sari, Risna and Rifa’i, Anggi Muhammad and Ahsan, Muhammad Salimy and Pahlevi, Mohammad Rezza and Arief, M Ilham},
  journal={International Journal of Research and Applied Technology (INJURATECH)},
  volume={2},
  number={2},
  pages={154--171},
  year={2022}
}

@inproceedings{yang2024effective,
  title={Effective distillation of table-based reasoning ability from llms},
  author={Yang, Bohao and Tang, Chen and Zhao, Kun and Xiao, Chenghao and Lin, Chenghua},
  booktitle={Proceedings of the 2024 Joint International Conference on Computational Linguistics, Language Resources and Evaluation (LREC-COLING 2024)},
  pages={5538--5550},
  year={2024}
}

@article{jiao_navigating_2025,
	title = {Navigating {LLM} ethics: advancements, challenges, and future directions},
	volume = {5},
	issn = {2730-5961},
	shorttitle = {Navigating {LLM} ethics},
	url = {https://doi.org/10.1007/s43681-025-00814-5},
	doi = {10.1007/s43681-025-00814-5},
	language = {en},
	number = {6},
	urldate = {2025-12-31},
	journal = {AI and Ethics},
	author = {Jiao, Junfeng and Afroogh, Saleh and Xu, Yiming and Phillips, Connor},
	month = dec,
	year = {2025},
	pages = {5795--5819},
}

@article{hinds_our_2023,
	title = {Our {Words} and the {Words} of {Artificial} {Intelligence}: {The} {Accountability} {Belongs} to {Us}},
	volume = {3},
	issn = {2691-3623},
	shorttitle = {Our {Words} and the {Words} of {Artificial} {Intelligence}},
	url = {https://journals.lww.com/cancercareresearchonline/fulltext/2023/04000/our_words_and_the_words_of_artificial.1.aspx},
	doi = {10.1097/CR9.0000000000000041},
	language = {en-US},
	number = {2},
	urldate = {2025-12-31},
	journal = {Cancer Care Research Online},
	author = {Hinds, Pamela S. and Bedinger Miller, Amy},
	month = apr,
	year = {2023},
	pages = {e041},
}

@article{kapania_im_2025,
	title = {'{I}'m {Categorizing} {LLM} as a {Productivity} {Tool}': {Examining} {Ethics} of {LLM} {Use} in {HCI} {Research} {Practices}},
	volume = {9},
	shorttitle = {'{I}'m {Categorizing} {LLM} as a {Productivity} {Tool}'},
	url = {https://doi.org/10.1145/3711000},
	doi = {10.1145/3711000},
	number = {2},
	urldate = {2025-12-31},
	journal = {Proc. ACM Hum.-Comput. Interact.},
	author = {Kapania, Shivani and Wang, Ruiyi and Li, Toby Jia-Jun and Li, Tianshi and Shen, Hong},
	month = may,
	year = {2025},
	pages = {CSCW102:1--CSCW102:26},
}

@inproceedings{ding_sustainable_2024,
	title = {Sustainable {LLM} {Serving}: {Environmental} {Implications}, {Challenges}, and {Opportunities} : {Invited} {Paper}},
	shorttitle = {Sustainable {LLM} {Serving}},
	url = {https://ieeexplore.ieee.org/abstract/document/10765824},
	doi = {10.1109/IGSC64514.2024.00016},
	urldate = {2025-12-31},
	booktitle = {2024 {IEEE} 15th {International} {Green} and {Sustainable} {Computing} {Conference} ({IGSC})},
	author = {Ding, Yi and Shi, Tianyao},
	month = nov,
	year = {2024},
	note = {ISSN: 2993-2084},
	pages = {37--38},
}

\end{document}